\documentclass[aps,prb,twocolumn,showpacs]{revtex4-1}
\usepackage{amsmath,amssymb}
\usepackage{graphicx}
\usepackage{verbatim}
\usepackage{amsfonts}
\usepackage{float}
\usepackage{mathrsfs}
\usepackage{dcolumn}
\usepackage{bm}
\usepackage{color}
\usepackage{ulem}
\usepackage[colorlinks,citecolor=blue]{hyperref}
\usepackage{wasysym}
\begin{document}

\newcommand{\odiff}[2]{\frac{\di #1}{\di #2}}
\newcommand{\pdiff}[2]{\frac{\partial #1}{\partial #2}}
\newcommand{\di}{\mathrm{d}}
\newcommand{\ii}{\mathrm{i}}
\renewcommand{\vec}[1]{{\mathbf #1}}
\newcommand{\vx}{{\bm x}}
\newcommand{\ket}[1]{|#1\rangle}
\newcommand{\bra}[1]{\langle#1|}
\newcommand{\pd}[2]{\langle#1|#2\rangle}
\newcommand{\tpd}[3]{\langle#1|#2|#3\rangle}
\renewcommand{\vr}{{\vec{r}}}
\newcommand{\vk}{{\vec{k}}}
\renewcommand{\ol}[1]{\overline{#1}}
\newtheorem{theorem}{Theorem}
\newcommand{\comments}[1]{}
\newcommand{\scrap}[1]{{\color{grey}{\sout{#1}}}}
\newcommand{\nts}[1]{[\emph{\color{red}{#1}}]}

\newcommand{\tj}[6]{ \begin{pmatrix}
  #1 & #2 & #3 \\
  #4 & #5 & #6
\end{pmatrix}}

\definecolor{darkgreen}{rgb}{0,0.5,0}
\definecolor{purple}{rgb}{0.35,0,0.35}
\definecolor{orange}{rgb}{1,0.5,0}
\definecolor{darkred}{rgb}{.7,0,0}
\definecolor{darkblue}{rgb}{0,0,.3}
\definecolor{grey}{rgb}{.6,.6,.6}
\definecolor{dimgreen}{rgb}{0.2,0.6,0.1}
\hypersetup{colorlinks,linkcolor=blue,urlcolor=blue,citecolor=blue}

\newcommand{\C}[1]{{\color{red}{#1}}}
\newcommand{\B}[1]{{\color{blue}{#1}}}
\newcommand{\Y}[1]{{\color{purple}{#1}}}
\newcommand{\jvd}[1]{{\color{magenta}{#1}}}
\newcommand{\changed}[1]{{\color{dimgreen}{#1}}}
\newcommand{\darkgreen}[1]{{\color{darkgreen}{#1}}}
\newcommand{\JVD}[1]{{\color{darkblue}{#1}}}
\newcommand{\jcomment}[1]{\textbf{\color{orange}[Comment JvD: {#1}]}}
\newcommand{\todo}[1]{\textbf{\color{red}[ToDo: {#1}]}}

\title{Hexagon-singlet solid ansatz for the spin-1 kagome antiferromagnet}

\author{Wei Li$^{1,2}$}
\author{Andreas Weichselbaum$^1$}
\author{Jan von Delft$^1$}
\author{Hong-Hao Tu$^{3}$}
\email{hong-hao.tu@mpq.mpg.de}
\address{1. Physics Department, Arnold Sommerfeld Center for Theoretical Physics, and Center for NanoScience, Ludwig-Maximilians-Universit\"at, 80333 Munich, Germany \\
2. Department of Physics, Beihang University, Beijing 100191, China \\
3. Max-Planck-Institut f\"ur Quantenoptik, Hans-Kopfermann-Str. 1, D-85748 Garching, Germany}

\begin{abstract}
  We perform a systematic investigation on the hexagon-singlet solid
  (HSS) states, which are a class of spin liquid candidates for the
  spin-1 kagome antiferromagnet. With the Schwinger boson
  representation, we show that all HSS states have exponentially
  decaying correlations and can be interpreted as a (special) subset
  of the resonating Affleck-Kennedy-Lieb-Tasaki (AKLT) loop states. We
  provide a compact tensor network representation of the HSS states,
  with which we are able to calculate physical observables
  efficiently. We find that the HSS states have vanishing topological
  entanglement entropy, suggesting the absence of intrinsic
  topological order. {We also employ the HSS states to perform a
  variational study of the spin-1 kagome Heisenberg antiferromagnetic
  model. When we use a restricted HSS ansatz preserving lattice
  symmetry, the best variational energy {per site} is found to be
  $e_0 = -1.3600$. In contrast, when allowing lattice symmetry breaking, we find a trimerized simplex 
  valence bond crystal with a lower energy, $e_0=-1.3871$.}
\end{abstract}
\date{\today}

\maketitle

\section{Introduction}
Frustrated antiferromagnets on the kagome lattice have attracted great
research interest recently. The ground state of a spin-1/2 kagome
Heisenberg antiferromagnet (KHAF) has been disclosed to be a
disordered state without any spontaneous symmetry breaking, i.e., a
quantum spin liquid. \cite{White-2011,Schollwoeck-2012} However, less
is known for the higher-spin KHAF models ($S>1/2$). Numerical studies
on the KHAF models with spin magnitude up to $S=3$ showed that,
\cite{Goetze-2011} while long-range magnetic order appears for
$S\geq3/2$, the ground states for the $S = 1/2$ and $S = 1$ KHAF
models remain non-magnetic. Experimentally, a number of spin-1 kagome
compounds have been synthesized and analyzed, e.g.,
m-MPYNN$\cdot$BF$_4$\cite{Awaga,Watanabe,Wada,Matsushita} and
Ni$_3$V$_2$O$_8$ \cite{Lawes}. The former has been found to be
non-magnetic even at very low temperatures (30mK),\cite{Watanabe} and
a spin gap has also been observed.\cite{Wada} This thus raises an
interesting question: do spin-1 kagome antiferromagnets also support
an intriguing spin liquid ground state?

\begin{figure}
\includegraphics[width=1\linewidth]{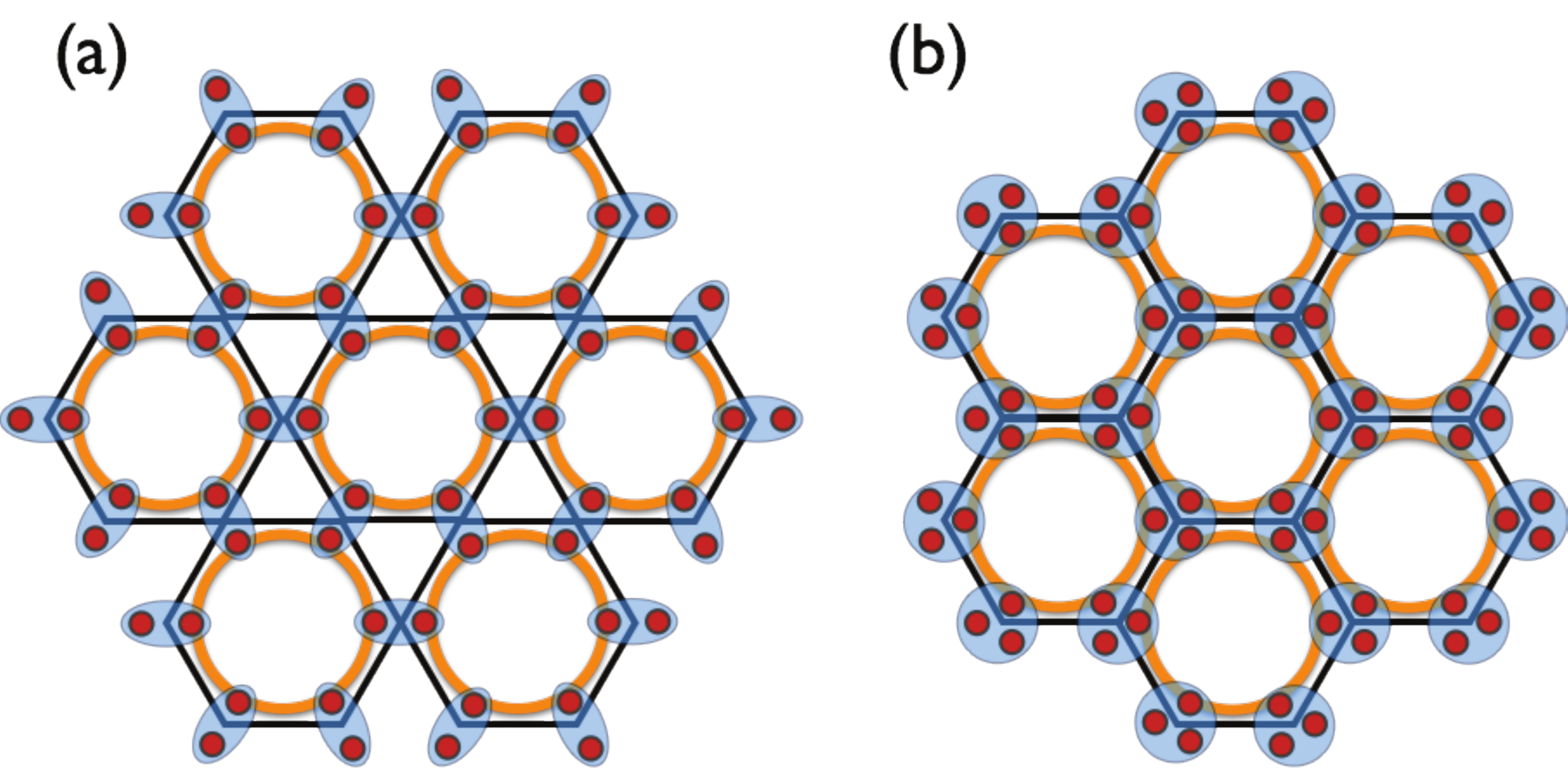}
\caption{(Color online) Schematic plots of (a) the spin-1 HSS ansatz
  on the kagome lattice and (b) the spin-3/2 HSS ansatz on the
  honeycomb lattices. \textit{Six} spin-1/2 virtual particles (red
  dots) form a spin singlet in each hexagon. In (a) the blue ovals
  denote projectors mapping two virtual spin-1/2's to a physical
  spin-1, while in (b) the circles denote projectors mapping three
  spin-1/2's onto a spin-3/2 space.}\label{fig:HSS}
\end{figure}

Various spin liquid proposals have been put forward for the spin-1
frustrated antiferromagnet.\cite{Liu-2010,Serbyn-2011,FWang-2011,Cenke-2012,Bieri-2012,Tu-2013,Hida-2000,HYao-2010,RAL-2014}
For the spin-1 KHAF model, Hida proposed a hexagon-singlet solid
(HSS) state as a candidate ground state. \cite{Hida-2000} This state
is constructed by projecting two virtual spin-1/2 particles around
each vertex of the kagome lattice into physical spin-1 degrees of
freedom. The name HSS refers to the fact that the six virtual
spin-1/2's in a hexagon of the kagome lattice form an entangled
singlet [see Fig. \ref{fig:HSS}(a)].  An alternative candidate is the
spin-1 resonating Affleck-Kennedy-Lieb-Tasaki loop (RAL)
state [see Fig. \ref{fig:ral}(b)].\cite{HYao-2010, RAL-2014} {The physical picture of the RAL state is the following: When representing the spin-1's in each site as two spin-1/2's (through symmetrization), the spin-1/2's from neighboring sites all form valence-bond singlets and, thus, each site has two valence bonds, which inevitably form closed AKLT loops \cite{Affleck-1987}. The RAL state is an \textit{equal} weight superposition of \textit{all} possible AKLT loops. On the kagome lattice, it belongs to a $\mathbb{Z}_2$ spin liquid. \cite{RAL-2014}} Compared to the HSS ansatz, the RAL has a lower variational energy (for the spin-1 KHAF model) on small clusters; while in the thermodynamic limit, the energy per site is clearly higher, $e_0\approx-1.27$. \cite{RAL-2014}

\begin{figure}
\includegraphics[width=0.5\linewidth]{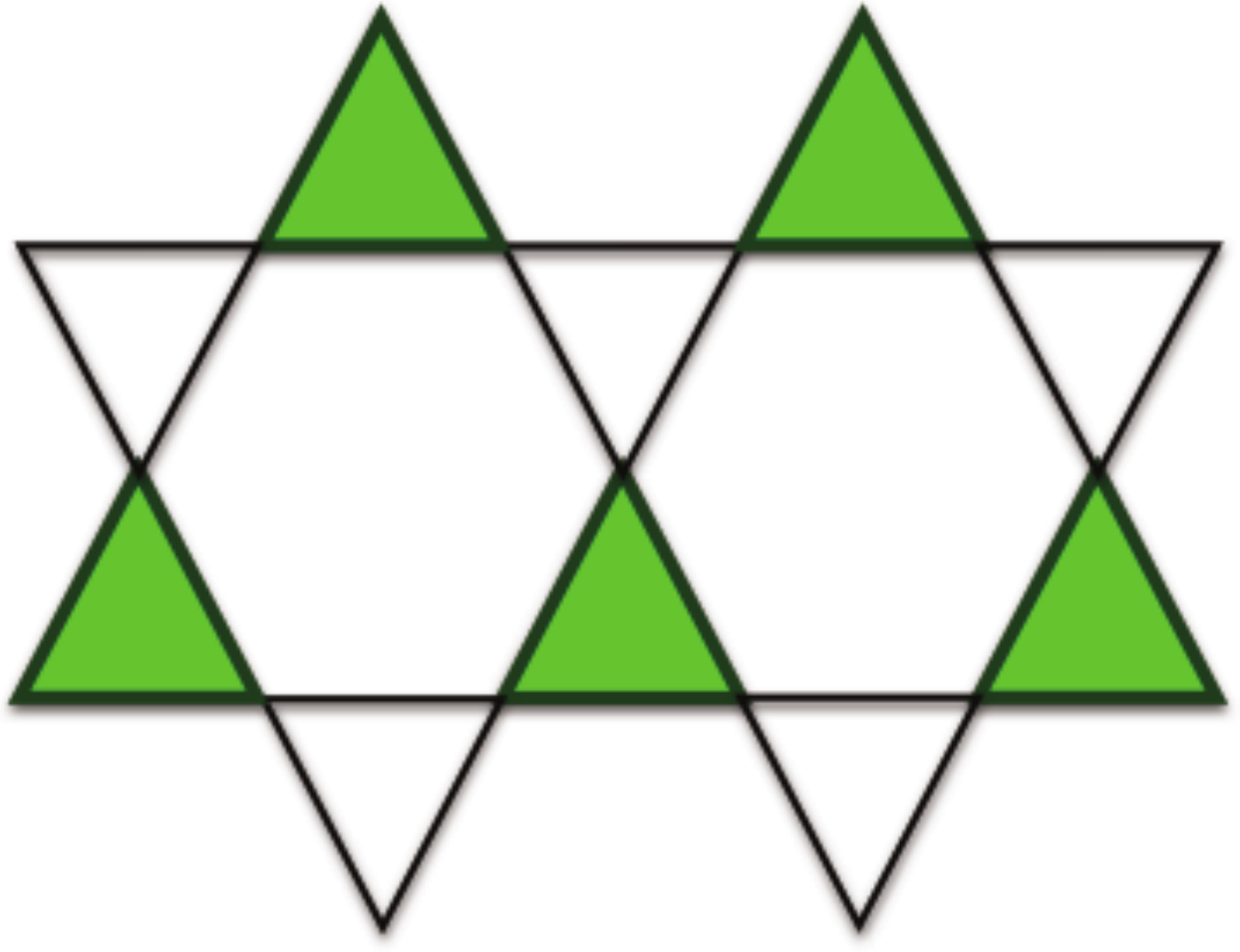}
\caption{(Color online) {Illustration of the spin-1 simplex valence bond crystal state on a kagome lattice. Two neighboring triangles have different energy expectation values, and the lattice inversion symmetry is broken.}}\label{fig:SVBC}
\end{figure}

Very recently, several extensive numerical studies have been devoted
to the spin-1 KHAF model, exploring its ground state
properties. Changlani and L\"auchli (CL) \cite{Lauchli-2014} employed
the density matrix renormalization group (DMRG)\cite{DMRG} to simulate
the model with cylindrical geometries; at the same time, Liu
\textit{et al.}\cite{Liu-2014} adopted tensor network
methods\cite{PEPS,Jordan-2008} to explore the same model on an
infinitely large kagome lattice and also on infinitely long cylinders
with various widths. These independent calculations, as well as
another related tensor network simulation by {Picot and
  Poilblanc},\cite{Poilblanc-2014} concluded that the ground state of
the spin-1 KHAF is non-magnetic, but that it breaks lattice inversion
symmetry and possesses a simplex valence-bond crystal (SVBC) order. 
{The simplex valence bond crystal is a non-magnetic state that favors trimerization,\cite{ZCai-2007} in that the energies (per triangle) differ between two neighoring triangles (see Fig. \ref{fig:SVBC}).} The energy per site of
this SVBC state was determined as $e_0 \approx -1.41$, in both DMRG
and tensor network
calculations.\cite{Lauchli-2014,Liu-2014,Poilblanc-2014}

{Even more recently, Nishimoto and Nakamura (NN14) \cite{Nishimoto-2014} came to a different conclusion: Based on DMRG calculations for clusters of various types of shapes and boundary conditions, they argued that the ground state of the spin-1 KHAF model is a HSS state and not the SVBC state advocated in Refs.~\onlinecite{Lauchli-2014,Liu-2014,Poilblanc-2014}. However, NN14 were not able to directly access states with HSS or SVBC structure; instead they sought to access them indirectly, using purposefully-designed boundary conditions that favor either HSS or SVBC structure. They then estimated the bulk values of $e_0$ by finite-size extrapolations to the thermodynamic limit. They reported $e_0 = -1.391(2)$ from SVBC-favoring clusters, $e_0 = -1.40988$ and $e_0 = -1.409(5)$ from clusters with cylindrical or periodic boundary conditions, respectively, and $e_0 = -1.41095$ from HSS-favoring clusters, thus concluding that HSS states win.

In our opinion, NN14's strategy is intrinsically flawed on very general grounds: in the thermodynamic limit boundary effects should vanish, thus a tool (here ground-state DMRG on finite-sized clusters) that \textit{relies on boundary effects} to distinguish two types of states (here HSS and SVBC), can not reliably estimate the difference in \textit{bulk} $e_0$ values for these two types of states. If different boundary choices lead to different finite-size extrapolated $e_0$ values, it just means that the clusters are not yet large enough to reliably capture the true bulk $e_0$ value of the true ground state, whatever it is. (More detailed comments on NN14's work are presented in Appendix~\ref{app:critique}.)

To reliably access the bulk properties of HSS-type states for the spin-1 KHAF, tools are needed that directly implement HSS structure in the variational candidate ground state, without relying on boundary effects.  In this work, we devise two such tools, one analytical, the other numerical, and use them to perform a systematic investigation of HSS states.

Our analytical tool is a formulation of the HSS ansatz in terms of SU(2) Schwinger bosons. In the Schwinger boson picture, we are able to {argue} that the HSS states have exponentially decaying correlation functions and thus describe gapped spin liquids. Furthermore, we also reveal that the HSS states have a hidden resonating AKLT-loop picture. However, we show that they have zero topological entanglement entropy,\cite{Kitaev-2006, Levin-2006} concluding that they are topologically trivial and thus do not belong to the same phase as the RAL state.

Our numerical tool is a compact tensor network representation of the HSS ansatz, with which we perform accurate tensor-network-based simulations. The weights of different hexagon-singlet configurations within a hexagon are treated as variational parameters to seek the lowest possible variational energy for the spin-1 KHAF model. It is found to be as low as $e_0=-1.3600$, which is {significantly} higher than the reported ground state energy $(-1.41)$ of the SVBC state.\cite{Lauchli-2014, Liu-2014} Moreover, we mimic a single step of imaginary-time evolution (on one of the two kinds of triangles), and thus add one additional parameter $\tau$. This gives a clear gain in energy ($e_0=-1.3871$), for a state breaking the lattice symmetry between two neighboring triangles. This variational calculation indicates that the HSS state, which does not break any lattice symmetry, may not be energetically favorable for the spin-1 KHAF model.}

The paper is organized as follows: in Sec. \ref{sec:hss} we introduce the HSS ansatz in terms of the Schwinger boson representation. In Secs. \ref{sec:tensor} and \ref{sec:variational} we show a compact tensor network representation for the HSS states, and use it to calculate the physical quantities, including the variational energy for the spin-1 KHAF model and various correlation functions. Sec. \ref{sec:conclusion} is devoted to the summary and discussions. 

\section{Hexagon-singlet solid ansatz}
\label{sec:hss}
In this section, we briefly review the hexagon-singlet solid ansatz. \cite {Hida-2000} We start in Sec. \ref{subsec:sb} by introducing the construction of the HSS ansatz in terms of Schwinger bosons, and then provide a physical picture of resonating AKLT loops for these states in Sec. \ref{subsec:ral}.

\subsection{Schwinger boson formulation}
\label{subsec:sb}
\begin{figure}[tbp]
\includegraphics[width=0.85\linewidth]{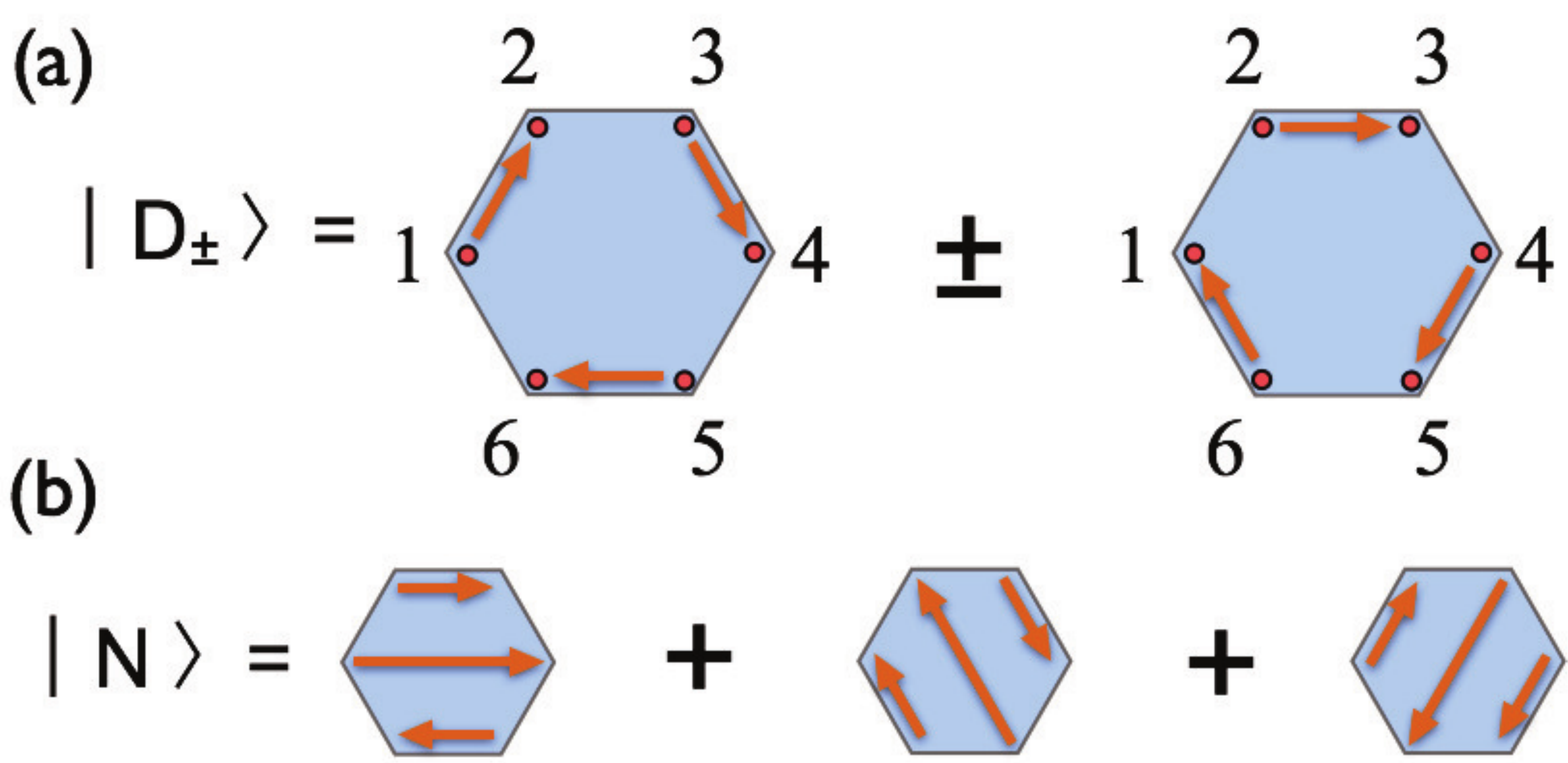}
\caption{(Color online) Graphical representation of the hexagon singlets. (a) All dimers are between nearest neighbors, like a resonating benzene ring, $\ket{D_{\pm}}$ is with $+$ ($-$) sign convention in the superposition. (b) An allowed dimer pattern $\ket{N}$ with \textit{longer-range} valence bonds in the hexagon. The arrow orientates from the first to the second spins in a singlet, which is anti-symmetric towards permutation of the two constituting spin-1/2's. Notice that these three hexagon singlets are all eigenstates of hexagon-inversion symmetry: $\ket{D_-}$ and $\ket{N}$ are odd (eigenvalue $-1$), while $\ket{D_+}$ is even ($+1$). {Moreover, $\ket{D_{\pm}}$ and $\ket{N}$ are all one-dimensional irreducible representations of the point group $C_{6v}$.}}
\label{fig:benzene}
\end{figure}

The hexagon-singlet solid ansatz for the spin-1 kagome antiferromagnet bears similarity to the construction of the 1D spin-1 AKLT
state.\cite{Affleck-1987} For each physical spin-1 site, one associates two virtual spin-1/2 particles. Since the kagome lattice
can be viewed as a lattice with corner-sharing hexagons, each hexagon contains six virtual spin-1/2 particles [see
Fig. \ref{fig:HSS}(a)]. On each hexagon, the six virtual spin-1/2 particles are combined into a SU(2) spin singlet. The final step is to
recover a physical spin-1 wave function by symmetrizing the two virtual spin-1/2's in the same lattice site into the spin-1
subspace. In contrast to ordinary AKLT states, there exist inequivalent singlet configurations per hexagon (depicted
schematically in Fig. \ref{fig:benzene}). Therefore, this construction provides a class of trial wave functions for the spin-1 kagome
antiferromagnet.

\begin{figure}[tbp]
\centering
\includegraphics[width=1\linewidth]{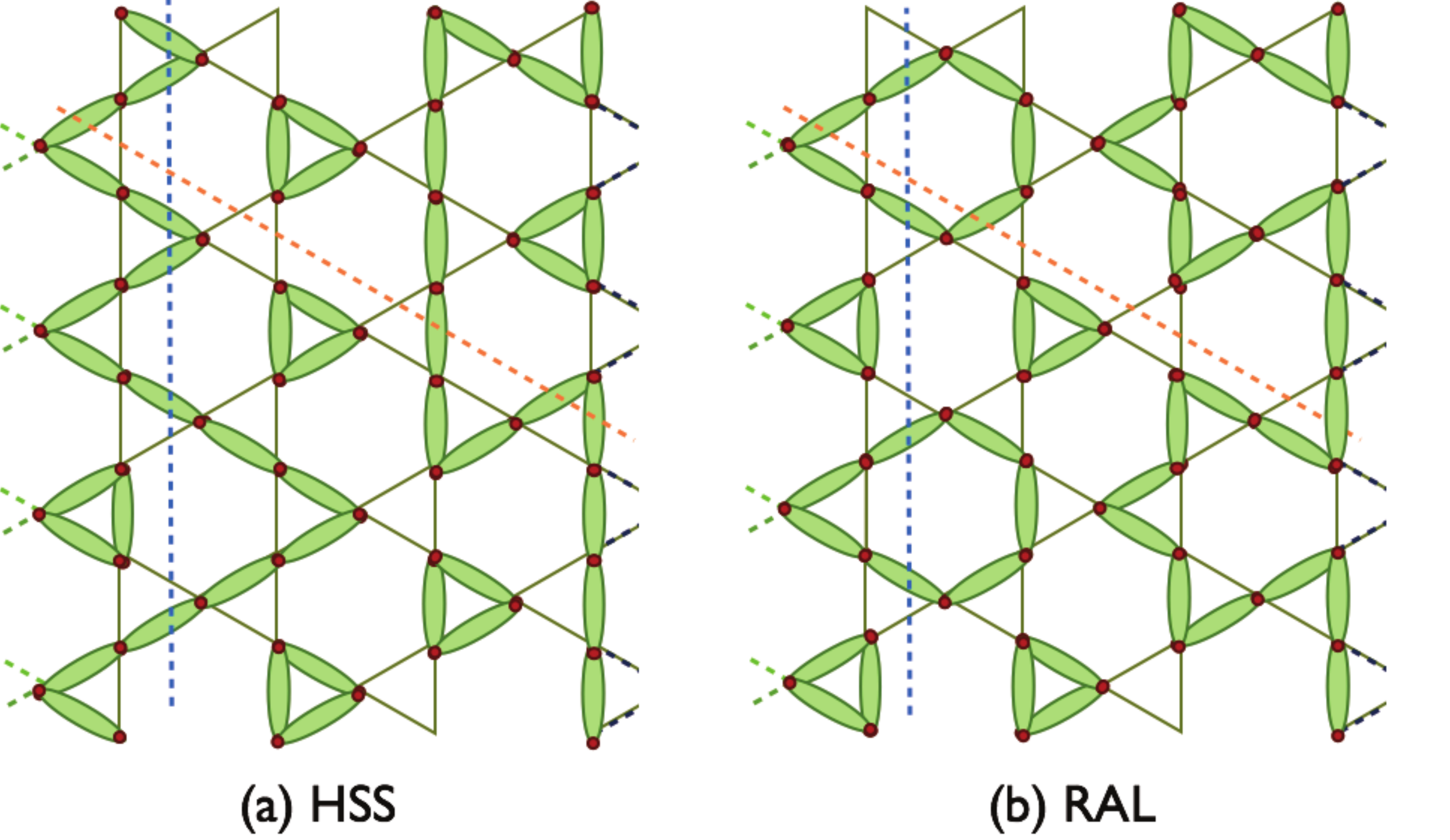}
\caption{(Color online) (a) Typical resonating AKLT-loop configuration arising from the benzene construction. {The green ellipses denote the valence bonds.} When cutting the loop configurations either vertically or at $60$ degrees to the vertical (denoted by dashed lines), it is only possible to intersect an even number of valence bonds. {(b) A typical RAL configuration, shown for comparison.} The RAL state has four virtual particles $0\oplus1/2$ on each vertex [see more details in Ref. \onlinecite{RAL-2014}], which is a fully packed equal weight superposition of \textit{all} possible loops, thus contains some configurations which are prohibited in the HSS state. {Periodic boundary conditions are assumed on both horizontal and vertical directions.}}
\label{fig:ral}
\end{figure}

Formally, it is convenient to formulate the HSS ansatz in terms of Schwinger bosons.\cite{Auerbach-book} In Schwinger boson language, the SU(2) spin operators for each site $i$ are represented as $S^a_i = \frac{1}{2} \sum_{\alpha \beta= \uparrow,\downarrow}
b^{\dag}_{i\alpha}\sigma^a_{\alpha \beta} b_{i\beta} \text{ \ } (a=x,y,z)$,
where $\sigma^a$ are Pauli matrices. On each site, a boson number constraint, $\sum_{\alpha=\uparrow,\downarrow}b^{\dag}_{i\alpha}b_{i\alpha}=2 $, has to be imposed to guarantee the physical spin-1 Hilbert space. Here the bosonic statistics takes care of the symmetrization of two virtual spin-1/2's into a spin-1 and, in terms of Schwinger bosons, the three spin-1 states are represented as
\begin{equation}
|1\rangle =\frac{(b_{\uparrow }^{\dagger })^{2}}{\sqrt{2}}|\mathrm{vac}
\rangle, \text{ \ }|0\rangle =b_{\uparrow }^{\dagger }b_{\downarrow
}^{\dagger }|\mathrm{vac}\rangle, \text{ \ }|-1\rangle =\frac{(b_{\downarrow
}^{\dagger })^{2}}{ \sqrt{2}}|\mathrm{vac}\rangle,
\end{equation}
where $|\mathrm{vac}\rangle$ is the vacuum of the Schwinger bosons.

By using the Schwinger bosons, the HSS state is written as
\begin{equation}
|\Psi \rangle = \prod_{\hexagon}\mathcal{P}_{\hexagon}^{+}| \mathrm{vac} \rangle,
\label{eq:HSS}
\end{equation}
where $\mathcal{P}_{\hexagon}^{+}$ creates a singlet state formed by six spin-1/2 Schwinger bosons within the same hexagon and can be generally written as
\begin{equation}
\mathcal{P}_{\hexagon}^{+} = \sum_{\alpha _{1}\alpha _{2}\cdots \alpha_{6}=\uparrow ,\downarrow }T_{\alpha _{1}\alpha _{2}\cdots \alpha _{6}}b_{\hexagon,\alpha _{1}}^{\dagger }b_{\hexagon,\alpha _{2}}^{\dagger }\cdots
b_{ \hexagon,\alpha _{6}}^{\dagger},
\label{eq:SingletCreation}
\end{equation}
{where $b^{\dagger}_{\hexagon, \alpha_i}$ are the Schwinger boson creation operators sitting at site $i$ of a hexagon [see Fig. 3(a)].} As there are several inequivalent ways of combining six spin-1/2's into a singlet, $T_{\alpha _{1}\alpha _{2}\cdots \alpha _{6}}$ in (\ref{eq:SingletCreation}) reflects this freedom of choice. For the wave function in (\ref{eq:HSS}), there are exactly two Schwinger bosons for every site and \textit{no extra projector} is needed to remove unphysical configurations.

To gain further insight into the HSS ansatz, we exploit the fact that the hexagonal singlet $\mathcal{P}_{\hexagon}^{+}|\mathrm{vac}\rangle $ in (\ref{eq:HSS}) can always be decomposed into a superposition of valence-bond singlets {(over-complete bases)} as
\begin{equation}
\mathcal{P}_{\hexagon}^{+}=\sum_{\{ij,kl,mn\}}w_{ij}w_{kl}w_{mn} \mathcal{C}_{\hexagon,ij}^{+} \mathcal{C}_{\hexagon,kl}^{+} \mathcal{C}_{\hexagon,mn}^{+},
\label{eq:HexagonalSinglet}
\end{equation}
where $\{ij,kl,mn\}$ denotes all allowed singlet-pair configurations (e.g., $ \{12,34,56\}$, $\{13,25,46\}$), the valence-bond singlet creation operator $\mathcal{C}$ is defined by
\begin{equation}
\mathcal{C}_{\hexagon,ij}^{+}=b_{\hexagon,i\uparrow }^{\dagger }b_{\hexagon,j\downarrow }^{\dagger
}-b_{\hexagon,i\downarrow }^{\dagger }b_{\hexagon,j\uparrow }^{\dagger },
\end{equation}
and $w_{ij}$ are coefficients controlling the weights of the valence bonds. Generically, $w_{ij}$ can be viewed as a set of free parameters. Comparing to (\ref{eq:SingletCreation}), an obvious advantage of the parametrization (\ref{eq:HexagonalSinglet}) is that the $C_6$ lattice symmetry can be easily imposed in these ansatz. For instance, one may consider a simple choice with $w_{ij}$ only depending on the distance between sites $i$ and $j$.

With the help of the Schwinger boson representation, we are able to {argue} that the HSS class of states have \textit{exponentially} decaying correlation functions, indicating that they describe a class of  \textit{gapped} spin liquids. The technical details on the proof of this statement are given in Appendix \ref{app:hss}. In short, the {argument} utilizes the spin-coherent state representation of the Schwinger boson states to write the norm of the HSS ansatz $\langle \Psi|\Psi \rangle$ as the partition function of a two dimensional (2D) classical statistical model describing interacting O(3) vectors on the kagome lattice. Additionally, the two-point correlation functions, say, the spin-spin correlation $\langle \Psi|\vec{S}_i \cdot \vec{S}_j|\Psi \rangle/\langle \Psi|\Psi \rangle $, can be expressed as the correlation function between O(3) vectors in the 2D statistical model. As the statistical model is at \textit{finite} temperature and has only short-range interactions, long-range order spontaneously breaking O(3) symmetry is not allowed, according to the Mermin-Wagner theorm,\cite{Mermin-1966} and the correlations between O(3) vectors (equivalently, spin-spin correlations in the HSS ansatz) decay exponentially. This is a direct generalization of the results in Ref. \onlinecite{Arovas-1988} showing that 2D AKLT states have exponentially decaying correlations.

In addition, the HSS construction is not restricted to the kagome lattice, but applies as well to any other lattices possessing hexagon motifs. For example, we show a spin-3/2 HSS ansatz on the honeycomb lattice in Fig. \ref{fig:HSS}(b), where the three spin-1/2 virtual particles surrounding a vertex are symmetrized to constitute the physical spin-3/2 degree of freedom. It is not difficult to see that our argument on the gapped spin liquid nature of the kagome HSS states (in Appendix \ref{app:hss}) also applies to all such HSS wave functions in 2D, including the spin-3/2 honeycomb HSS state in Fig. \ref{fig:HSS}(b).

\subsection{Resonating Affleck-Kennedy-Lieb-Tasaki loop picture}
\label{subsec:ral}
Let us now introduce a simple example belonging to the HSS class on the kagome lattice, which we call the Benzene Ring State (BRS). Based on the BRS example, we uncover a resonating AKLT loop picture for the HSS ansatz.

The BRS states are defined by restricting hexagonal singlets in (\ref{eq:HexagonalSinglet}) to short-range dimers between \textit{neighboring} sites. Two (inequivalent) such choices for $w_{ij}$ in (\ref{eq:HexagonalSinglet}) are given by
\begin{equation}
w_{12}=w_{34}=w_{56}=1,\text{ \ }w_{23}=w_{45}=w_{61}=1,
\label{eq:benzene1}
\end{equation}
and
\begin{equation}
w_{12}=w_{34}=w_{56}=1,\text{ \ }w_{23}=w_{45}=w_{61}=-1,
\label{eq:benzene2}
\end{equation}
respectively. The graphical representations for the above two hexagonal singlet choices, resembling resonating benzene rings, are shown in Fig. \ref{fig:benzene}(a).

When building the wave function (\ref{eq:HSS}) using (\ref{eq:benzene1}) or (\ref{eq:benzene2}), expanding the product
$\prod_{\hexagon}$ in (\ref{eq:HSS}) leads to a number of \textit{nearest-neighbor} valence-bond configurations. A typical
configuration is shown in Fig. \ref{fig:ral}(a). An interesting observation is that this configuration can be viewed as the covering
of spin-1 AKLT loops on the kagome lattice. This is due to the fact that each site shares two spin-1/2 valence bonds (forming spin-1
physical sites), that is to say, every site is involved in two valence bonds, which inevitably form fully packed loop
structure. Based on this observation, we conclude that the BRS state can be viewed as an \textit{equal} weight superposition of resonating AKLT loops.

What about the HSS ansatz with longer-range valence bonds within each hexagon? It is not difficult to see that, when expanding the product $\prod_{\hexagon}$ in (\ref{eq:HexagonalSinglet}), the AKLT loop structure in each configuration is still preserved, though the loops can connect sites beyond nearest neighbors (NN). Then, the role of the weights $w_{ij}$ in (\ref{eq:HexagonalSinglet}) is to control the loop tension. This shows that \textit{all} HSS ansatz (\ref{eq:HSS}) belong to the broader family of resonating AKLT loop states. 

{To be concrete, we consider the following expansion of Hida's hexagon singlet (note that the valence bond basis states are non-orthogonal and over-complete, thus the expansion below is not the unique choice, see more discussions in Appendix \ref{variational}):
\begin{equation}
\ket{ G} = \ket{\rm{D_-}}+ \omega \ket{N}
\label{eq:GDN}
\end{equation}
where $\ket{\rm{D_-}}$ and $\ket{N}$ are illustrated in Fig. \ref{fig:benzene}, and $|G\rangle$ is the ground state (a hexagon singlet) of a six-site NN Heisenberg ring model, firstly used by Hida in constructing his HSS state. A straightforward calculation reveals that $\omega \simeq 0.5826$ (also seen in Fig. \ref{fig:EsHSS}). Therefore, the corresponding $w_{i,j}$ coefficients of the HSS state read:
\begin{eqnarray}
&& w_{12}= w_{34} = w_{56}= 1 + \omega, \notag \\
&& w_{23} = w_{45}=w_{61}= -1 + \omega, \notag \\
&& w_{14}= w_{52}=w_{36}= \omega,
\end{eqnarray}
which clearly demonstrates the resonating AKLT-loop nature of Hida's HSS state.}

Coming back to the BRS states which can be regarded as an equal weight superposition of AKLT loops, it is rather interesting to make a comparison between them with the RAL state considered in Ref. \onlinecite{RAL-2014}. In the latter, the RAL state is an equal weight superposition of \textit{all} possible AKLT loops [see Fig. \ref{fig:ral}(b)] and has $\mathbb{Z}_2$ topological order on the kagome lattice. However, {there is an additional constraint in the BRS due to the benzene ring construction which requires that} no loop can be formed that contains two successive valence bonds within the same hexagon. This means that the allowed loop configurations in the BRS states are strictly less than those of the RAL state. Actually, this leads to an important observation which reveals the difference between them: When the BRS state is defined on a torus, cutting the torus in either horizontal or vertical direction always intersects an\textit{even} number of valence bonds [see Fig. \ref{fig:ral}(a)]. However, this is not the case for RAL states, where such a cut can intersect an even or odd number of valence bonds [Fig. \ref{fig:ral}(b)]. {To be precise, there are four types of RAL states corresponding to the four combinations of parities for the number of valence bonds encountered along horizontal cuts (even/odd) and vertical cuts (even/odd). For a given RAL state, the parities are invariant when the cuts are swept through the lattice.}\cite{RAL-2014} While the existence of such ``parity'' sectors is essential for $\mathbb{Z}_2$ topological order, this already gives a hint that there is no topological order (at least not $\mathbb{Z}_2$ type) in the BRS state and they are distinct from the RAL state, even though they share very similar loop structure with the latter. In Sec. \ref{sec:tensor}, we will provide numerical evidence that the HSS ansatz, including the BRS state, has vanishing topological entanglement entropy and thus no topological order.

\begin{figure}
\includegraphics[width=1\linewidth]{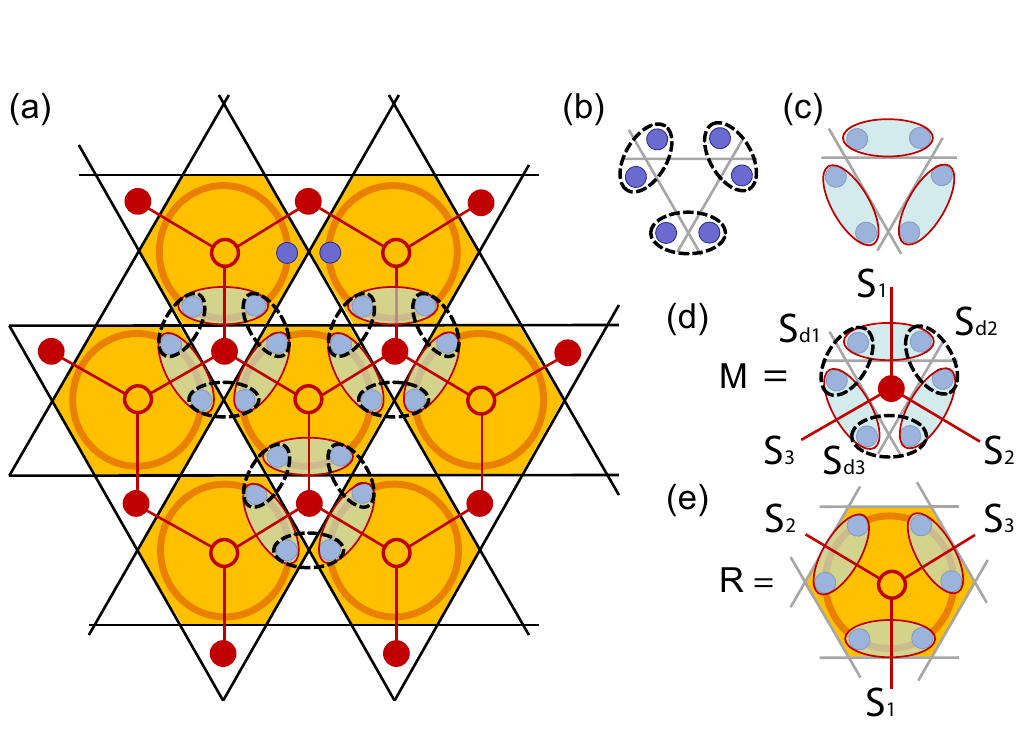}
\caption{(Color online) (a) Tensor network representation of the HSS ansatz on the kagome lattice. (b) The local projector $P$ (dashed ovals) maps two spin-1/2 virtual particles onto the physical spin-1 space. (c) By fusing two virtual spins, we thus introduce composite virtual particle $0\oplus 1$ on the bond. (d) A projection tensor {$M_{(S_1,m_1) (S_2, m_2) (S_3, m_3)}^{(S_{d_1}, m_{d_1}) (S_{d_2}, m_{d_2}) (S_{d_3}, m_{d_3})}$ [$(S_i,m_i)$'s are geometric indices along red solid lines and $(S_{d_i}, m_{d_i})$'s are physical indices], and (e) a rank-three hexagon tensor $R_{(s_1,m_1) (s_2, m_2) (s_3, m_3)}$} are obtained using the composite virtual particles. {$S$ and $m$ denote the spin and magnetic quantum numbers, respectively.}}
\label{fig:tensor} 
\end{figure}

\begin{figure}
\includegraphics[width=0.95\linewidth]{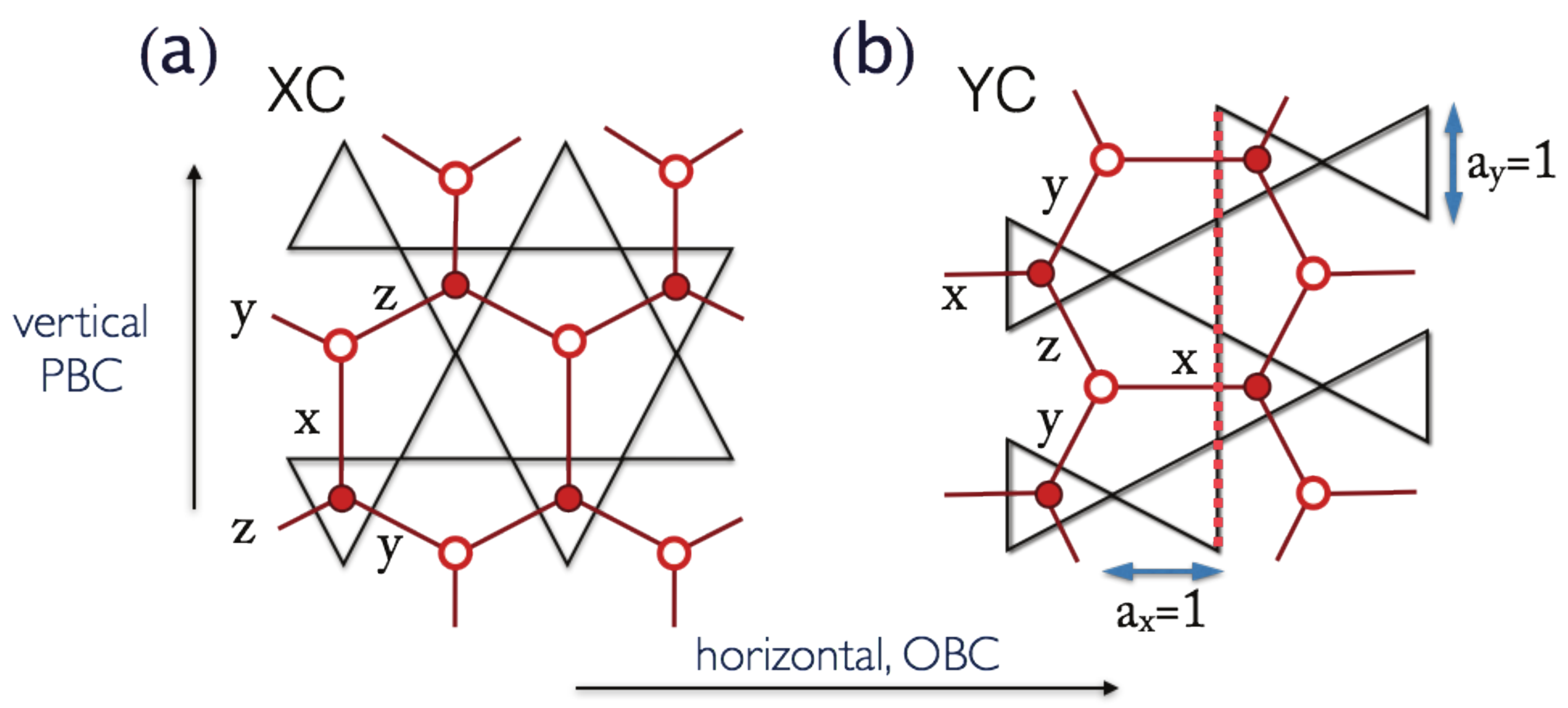}
\caption{(Color online) The tensor networks on (a) XC and (b) YC cylindrical geometries, where periodic (open) boundary condition in vertical (horizontal) direction is assumed. The length units $a_{x,y}=1$ are also shown.}
\label{fig:XCYC}
\end{figure}

\section{Tensor network representation and simulation of the hexagon-singlet solid states}
\label{sec:tensor}
In this section, we provide a compact tensor network representation for the HSS states (see Fig.~\ref{fig:tensor}) and calculate the physical quantities by using tensor-network-based simulations. In Sec. \ref{subsec:brs} we start with analyzing the BRS, a special HSS state introduced in Sec. \ref{subsec:ral}. In Sec. \ref{subsec:hss} we move on to the study of another special HSS state introduced in Ref. \onlinecite{Hida-2000}.

According to Sec. \ref{sec:hss}, the HSS state (\ref{eq:HSS}), instead of its Schwinger boson formulation, can be alternatively written as
\begin{equation}
|\Psi\rangle = \bigotimes_i P_i \prod_{\hexagon} | \psi_{\hexagon} \rangle,
\label{eq:psi}
\end{equation}
and
\begin{equation}
|\psi_{\hexagon} \rangle = \sum_{\{\sigma \in \hexagon\}} T_{\sigma_i \sigma_j \sigma_k \sigma_l \sigma_m \sigma_n} |\sigma_i, \sigma_j, \sigma_k, \sigma_l, \sigma_m,\sigma_n \rangle,
\label{eq:psi_hexagon}
\end{equation}
where $\sigma_i$ is a virtual spin-1/2 located at site $i$ belonging to the hexagon. $|\psi_{\hexagon}\rangle$ denotes the hexagon singlet formed by six virtual spin-1/2 particles, and $P$ projects two virtual particles onto the triplet subspace
\begin{equation}
P= \sum_{\sigma_1, \sigma_2} \sum_{m}  C_{\sigma_1, \sigma_2}^m |m\rangle \langle \sigma_1,\sigma_2|,
\label{eq:projector}
\end{equation}
where $m \in \{\pm1,0\}$ denotes the physical, $\sigma_1,\sigma_2 \in \{\pm1/2\}$ the virtual space, and $C_{\sigma_1, \sigma_2}^m$ is the Clebsch-Gordan coefficient symmetrizing two spin-1/2's into a physical spin-1, with nonvanishing elements as $C^{1}_{1/2,1/2} = C^{-1}_{-1/2,-1/2}=1$ and $C^{0}_{1/2,-1/2} = C^{0}_{-1/2,1/2}=1/\sqrt{2}$. In Fig. \ref{fig:tensor}(a), the HSS structure is depicted: the sixth-order tensor $T$ is represented by the ring within each hexagon and the dashed oval on each site indicates the projector $P$ [Fig. \ref{fig:tensor}(b)]. 

However, from a numerical point of view, this representation is not practical for calculations, owing to the high coordination number ($z=6$) of the $T$-tensors. To overcome this difficulty, we here introduce a scheme shown in Figs.~\ref{fig:tensor}(d,e): two neighboring spin-1/2 particles (in the hexagon) are fused into a composite virtual particle $0 \oplus 1$ of dimension four, and the coordination number of  the hexagon tensor is lowered down to $z=3$. After this transformation, we get a hexagonal tensor network consisting of $M$ and $R$ tensors, which can be more easily treated with the tensor network techniques. Notice that one has the freedom to block the virtual particles in two different ways [odd-even, or even-odd pairs, as in Fig. \ref{fig:benzene}(a)], they represent essentially the same state.

\subsection{The Benzene Ring State}
\label{subsec:brs}

We start with the BRS depicted in Fig.~\ref{fig:ral}, which only contains NN valence bonds. There are two hexagon-singlet
configurations in the BRS for two sign choices, i.e., $\ket{\psi_{\hexagon}}$ in Eq.~(\ref{eq:psi}) can be chosen as
$\ket{D_{\pm}}$ in Fig.~\ref{fig:benzene}(a), and we thus construct $\ket{\rm{BRS}}_{E,O}$ $= \bigotimes_i P_i \prod_{\hexagon} \ket{D_{\pm}}$. {The details of tensors $M$ and $R$ which constitute the SU(2)-invariant tensor network representation of BRS can be found in Appendix \ref{QSpace-Tensor}}. 

We take the BRS as a variational ground-state wave function of the spin-1 KHAF model (with the Hamiltonian $H_{\rm{KHAF}}=\sum_{\langle i,j \rangle} \vec{S}_i \cdot \vec{S}_j$) and first calculate the energy per site $e_0$ using the infinite projected entangled-pair state (iPEPS) contraction algorithm [via the boundary matrix product state (MPS) scheme].\cite{PEPS, Jordan-2008} The resulting energy per site is $e_0=-1.31670602$ for $|\rm{BRS}\rangle_O$, while $e_0=-0.831271138$ for $|\rm{BRS}\rangle_E$.

\begin{figure}
\includegraphics[width=1\linewidth]{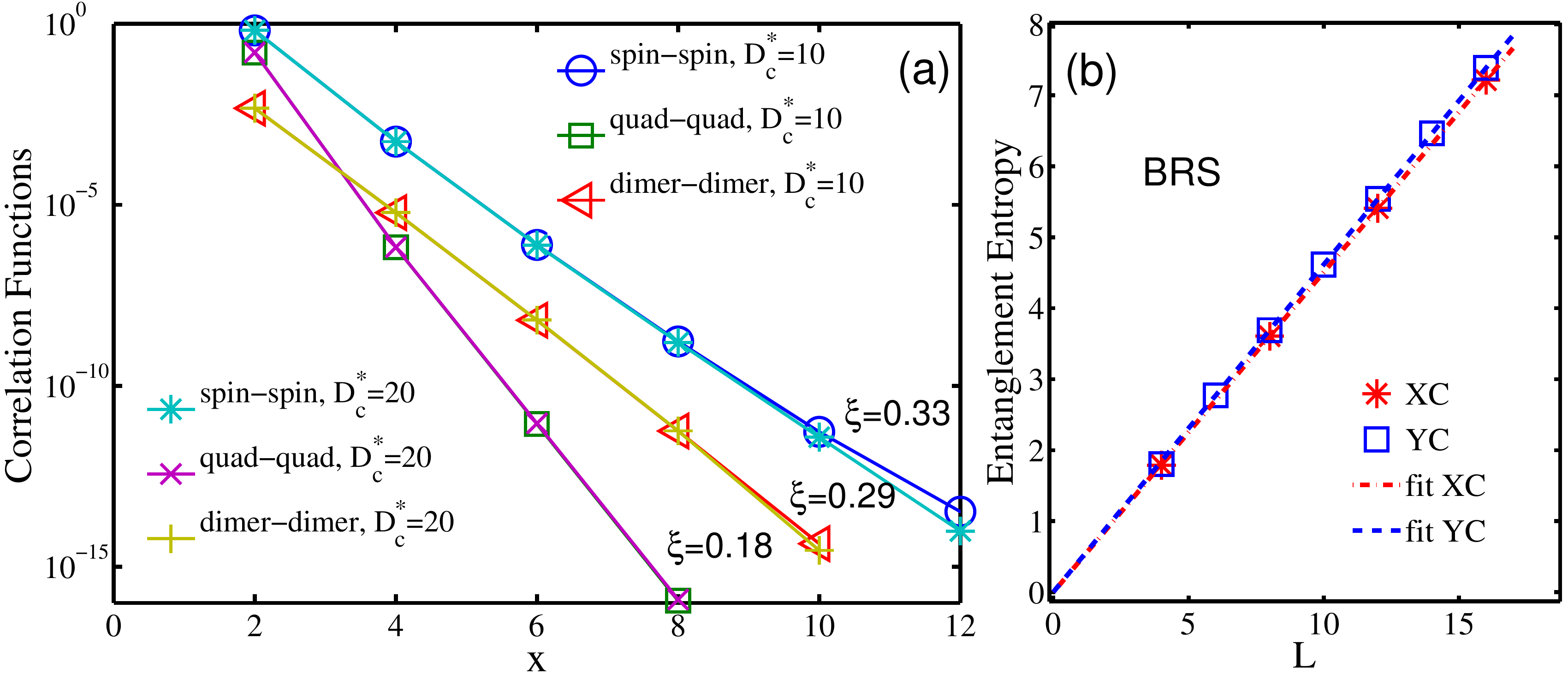}
\caption{(Color online) (a) The spin-spin, quadrupole-quadrupole, and dimer-dimer correlation functions of the $\ket{\rm{BRS}}_O$, obtained by SU(2) iPEPS contractions. All of the correlation functions $C(x)$ are found to decay exponentially, as $C(x) \sim \exp(-x/\xi)$. The correlation lengths $\xi$ are obtained from fitting the data. (b) Entanglement entropies of the odd BRS on XC and YC geometries with various circumferences (up to $L=16$), the entropy data extrapolate to $\gamma=0$ in the $L=0$ limit.}
\label{fig:CF}
\end{figure}

In Fig. \ref{fig:CF}(a), we show the numerical results of various correlation functions of $\ket{\rm{BRS}}_O$, which are also obtained by iPEPS contractions (thus measured on an infinite kagome lattice). The correlation functions include the spin-spin $C_{SS}(j-i) = \langle \vec{S}_i \cdot \vec{S}_j \rangle$, the quadrupole-quadrupole $C_{QQ}(j-i)=\langle \vec{Q}_i \cdot \vec{Q}_j \rangle$, and the dimer-dimer $C_{DD}(j-i)=\langle (\vec{S}_i \cdot \vec{S}_{i+1}) (\vec{S}_j \cdot \vec{S}_{j+1}) \rangle - \langle \vec{S}_i \cdot \vec{S}_{i+1} \rangle  \langle \vec{S}_j \cdot \vec{S}_{j+1} \rangle$ correlations. All the correlation functions are calculated in an SU(2)-invariant manner, i.e., the $C_{SS}$ and $C_{QQ}$ correlations are computed using irreducible tensor operators $\vec{S}^{(1)} = \{S^+, S^z, S^-\}$ and $\vec{Q}^{(2)}=\{(S^+)^2,-(S^+S^z+S^zS^+), \sqrt{\frac{2}{3}}(3(S^z)^2-2), (S^-S^z+S^zS^-), (S^-)^2\}$, respectively. The correlations are measured along the vertical line marked as a red dashed line in Fig. \ref{fig:XCYC}(b), with length unit $a_y=1$ being specified there, and are found to decay exponentially, as expected from the rigorous proof in terms of Schwinger bosons (Sec. \ref{sec:hss} and Appendix \ref{app:hss}). The correlation lengths $\xi$, extracted by linear fittings from the semi-log plot, are found to be rather short.

Besides the infinite kagome lattice, we are also interested in evaluating the properties of BRS on the cylindrical geometries (see the XC and YC geometries in Fig. \ref{fig:XCYC}). In Fig. \ref{fig:CF}(b), we show the entanglement entropy results {[$S(L)=-\rm{tr}(\rho\ln\rho)$, where $\rho$ is the half-cylinder reduced density matrix]} of XC and YC geometries, versus various cylinder circumferences (up to $L=16$ for both geometries), which measures the quantum entanglement between two half-infinite cylinders. As shown in Fig. \ref{fig:CF}(b), we extrapolate $S(L)$ using the formula $S=\alpha L - \gamma$,\cite{Jiang-2012} and get $\gamma\simeq0$ as $L \to 0$, in both XC and YC cases. This observation shows unambiguously that the BRS possesses no long-range entanglement and thus no intrinsic topological order, this is due to the local constraint arising from the benzene construction (related discussions in Sec. \ref{subsec:ral}).

\begin{figure}
\includegraphics[width=1\linewidth]{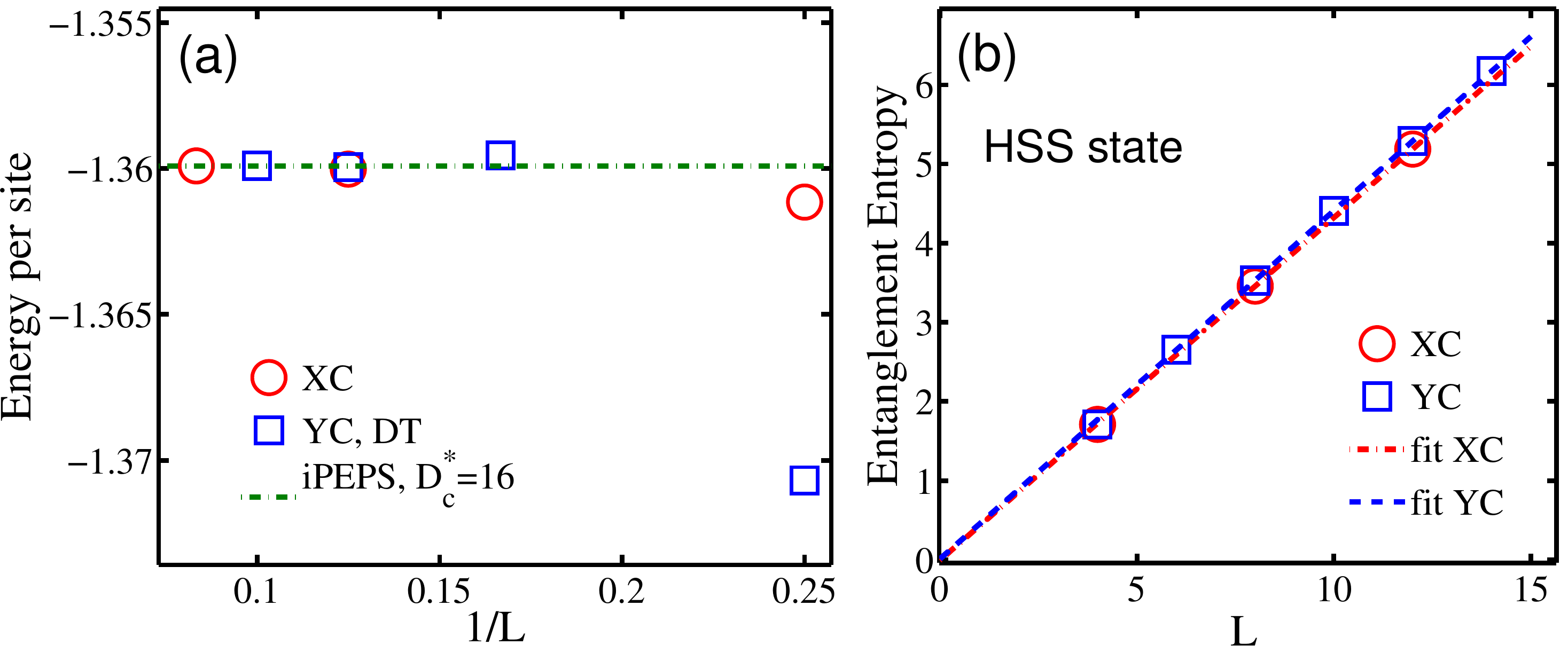}
\caption{(Color online) (a) Energy expectation values (per site) $e_0$ of Hida's HSS state on cylinders with various circumferences $L$ [see Fig. \ref{fig:XCYC} for the illustration of the XC and YC geometries, the length units $a_x,a_y=1$ are also shown in Fig. \ref{fig:XCYC}(b)]. The energy results converge very fast to the thermodynamic limit obtained by iPEPS contractions. (b) Entanglement entropy results extrapolate to $\gamma\simeq0$ as $L\to 0$, for both XC and YC geometries.}
\label{fig:cylinder}
\end{figure}

\subsection{Hida's Hexagon-Singlet Solid State}
\label{subsec:hss}
Next, we turn to the original HSS state proposed in Ref.~\onlinecite{Hida-2000} (henceforth to be referred to as Hida's HSS state, or more briefly the Hida state), where the hexagon singlet $|\psi_{\hexagon} \rangle$ in Eq. (\ref{eq:psi}) is chosen to be the ground state of a six-site (hexagonal) spin-1/2 Heisenberg ring,
\begin{equation}
H_{\hexagon}=J_1 \sum_{\langle i,j \rangle} \vec{S}_i \cdot \vec{S}_j + J_2 \sum_{\langle\langle i,j \rangle\rangle} \vec{S}_i \cdot\vec{S}_j + J_3 \sum_{\langle\langle\langle i,j \rangle\rangle\rangle} \vec{S}_i\cdot \vec{S}_j,
\label{eq:hexagon}
\end{equation}
where $J_1$ is the NN coupling, $J_2, J_3$ are the second- and third-NN interactions, and $\vec{S}$'s are the $S=1/2$ spin operators. Following {the original definition of Hida's HSS state in} Ref.~\onlinecite{Hida-2000}, we also diagonalize the Heisenberg ring with only NN couplings (couplings $J_1=1, J_2=J_3=0$) in a hexagon, and find five orthonormal singlet eigenstates, with energies $-2.8028$,  $-1.5000$, $-0.5000$, $-0.5000$, and $0.8028$, with a considerable gap ($\sim 1.3$) between the resulting ground and the first excited states. {After fixing the ground-state hexagon singlet, we can obtain the tensor network representation ($M$ and $R$ tensors) of the Hida state (see Appendix \ref{QSpace-Tensor} for more details). }

\begin{table}[h]
\centering
\caption{Energy expectation values of Hida's HSS state for the spin-1 KHAF model, obtained by SU(2) iPEPS contractions. {$D_c^*$ ($D_c$) is the number of multiplets (individual states) retained on the MPS bond. We show thirteen significant digits for $e_0$, since $e_0$ is converged to that accuracy upon retaining $D_c^*$ larger than 10.}}\label{tab:Es}
\begin{tabular}{ c c c c c }
\hline
  $D_c^*$ & $D_c$ & $e_0$ & truncation error \\
  4 & 8 & $-1.359944730698$ & $2\times 10^{-6}$ \\
  5 & 11 & $-1.359910140148$ & $4\times 10^{-8}$ \\
  6 & 16 & $-1.359909517302$ & $2\times 10^{-12}$ \\
  10 & 26 & $-1.359909517316$ & $2\times 10^{-13}$ \\
  16 & 44 & $-1.359909517316$ & $2\times 10^{-15}$ \\
\hline
\end{tabular}
\end{table}

We again consider two kinds of geometries for the evaluation of observables: the infinite kagome lattice and cylinders (including XC and YC geometries). Firstly, the energy per site $e_0$ is calculated through iPEPS contractions and the results are shown in Tab. \ref{tab:Es}. The small truncation error suggests that the data are very well converged when more than $D_c^*=10$ multiplets (corresponding to $D_c=26$ states) are retained in the geometric bond of the boundary MPS.

Besides the iPEPS calculations, we also performed exact contractions on various cylinders, the results are shown in Fig.~\ref{fig:cylinder}(a). Notably, the energy expectation value is determined as $e_0=-1.359910231678$ for XC12, in excellent agreement with the accurate iPEPS results in Tab. \ref{tab:Es}. This value is also in accordance with that in Ref. \onlinecite{Lauchli-2014}, where the HSS energy is estimated as $-1.36$ based on the exact diagonalization results on several small clusters. This variational energy is lower than that of the RAL state ($\approx -1.27$) in Ref. \onlinecite{RAL-2014}, but still higher than the best estimate $e_0\simeq-1.410$ (of an SVBC state) in Refs. \onlinecite{Lauchli-2014, Liu-2014} for the actual ground state of the spin-1 KHAF model.

In Fig.~\ref{fig:cylinder}(b), we show the entanglement entropies on (both XC and YC) cylinders of various circumferences $L$. They extrapolate to zero in the $L=0$ limit, meaning that Hida's HSS state possesses no intrinsic topological order, just as the $\ket{\rm{BRS}}_O$ investigated in Sec.~\ref{subsec:brs}. This is an expected and consistent observation, because the rigorous proof in Sec.~\ref{sec:hss} guarantees that all HSS states are gapped spin liquids and thus the Hida state should belong to the same (non-topological) phase as the BRS.

In addition, we also studied the entanglement spectra (ES) of the HSS states on various cylinders through exact contractions.\cite{Cirac-2011, Poilblanc-2013} In the Appendix \ref{app:spectra}, we show results at the Hida point, where a nonvanishing triplet gap has been observed in the ES [Fig. \ref{fig:ES}(a)]. This is in contrast to the $S=2$ AKLT state on a square lattice, where the gaps in the ES decrease as the system size increases, and finally vanish in the thermodynamic limit [see Fig. \ref{fig:ES}(b)].\cite{Cirac-2011, Lou-2011} The absence of a gapless edge excitation  in the HSS state indicates that it has no symmetry-protected topological (SPT) order, since gapless edge modes necessarily appear in the SPT phases.\cite{XieChen,Zheng-Xin}

\section{Variational study of the kagome Heisenberg antiferromagnetic model}
\label{sec:variational}
In this section, we discuss variational energies of the HSS states for the spin-1 KHAF model, and furthermore search for lower variational energy within the present tensor network ansatz with $D^*$=2 {(i.e., two multiplets $0\oplus 1$, which contain $D=4$ individual states)}. Among various ways to perform the variational studies, we will discuss three cases below, which turn out to produce consistent results.

\begin{figure}
\includegraphics[width=1\linewidth]{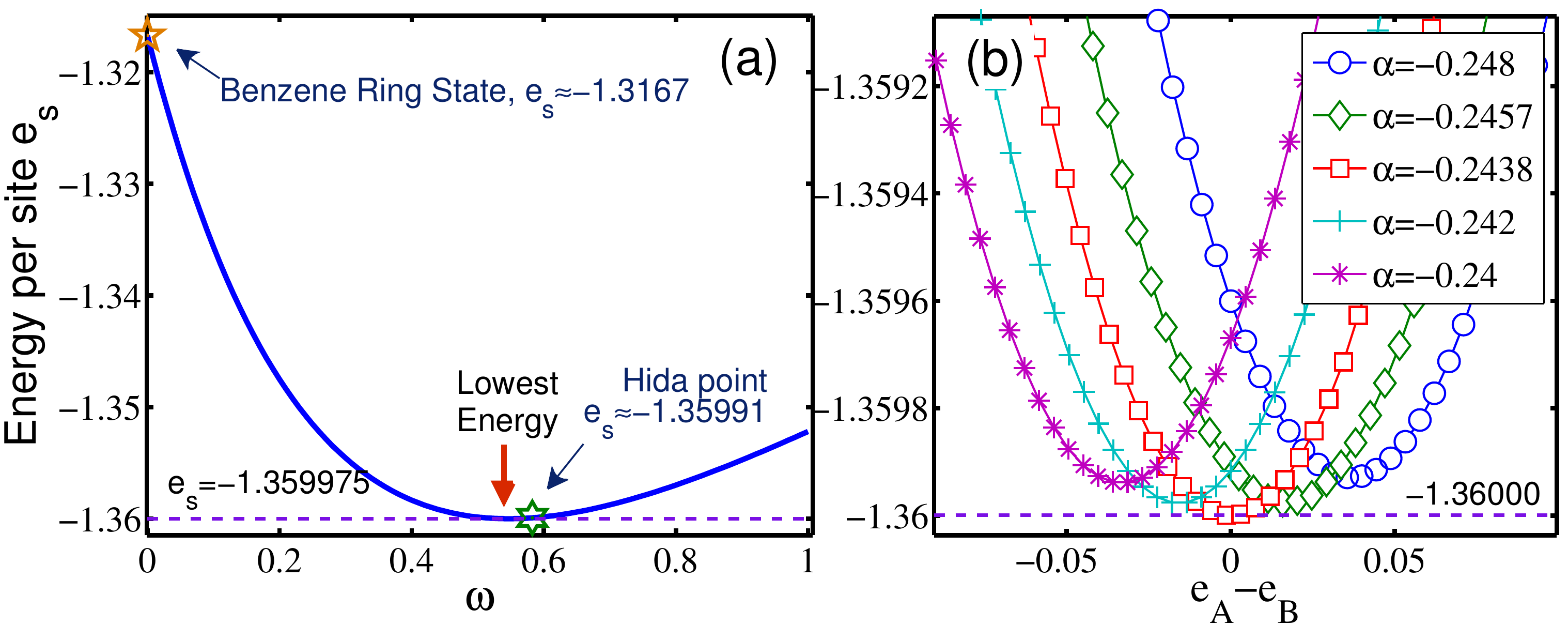}
\caption{(Color online) Variational energies of the HSS states. (a) By adding the hexagon-singlet configurations in Fig.~\ref{fig:benzene}(b), with weight $\omega$, to the benzene ring state, we connect smoothly the latter ($\omega=0$) with the Hida point ($\omega=0.582618977$). The lowest variational energy ($e_0=-1.359975$) is slightly lower than that of the Hida point. (b) By tuning parameters $\alpha$ and $\beta$, we obtain the lowest energy $e_0 \simeq -1.36000$ with $\alpha=-0.2438$ and $\beta=0.134$, which has no energy difference between the two kinds of triangles, i.e., $e_A\simeq e_B$.}
\label{fig:EsHSS}
\end{figure}

\begin{figure}
\includegraphics[width=0.85\linewidth]{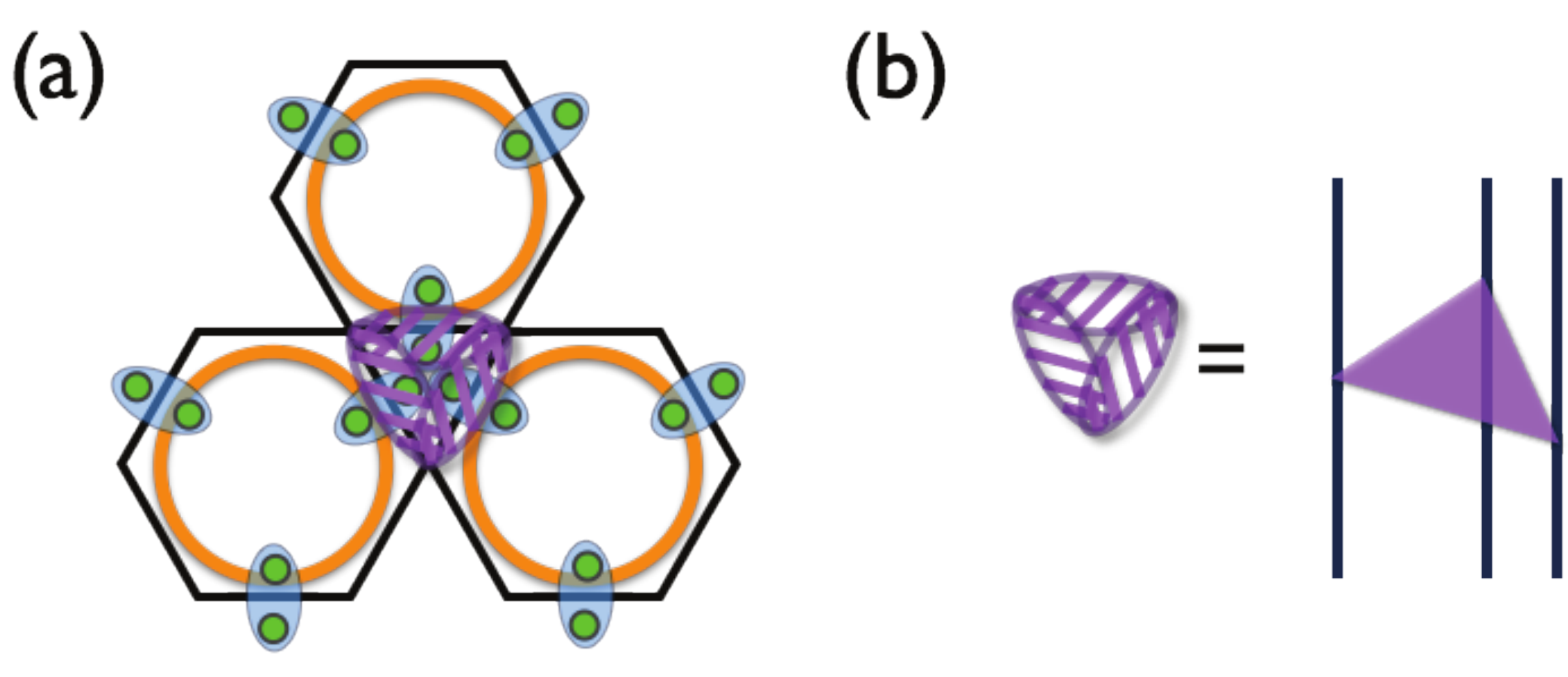}
\caption{(Color online) Variational study of the spin-1 KHAF model beyond the HSS ansatz. (a) Although the hexagon singlets are intact, applying an imaginary-time evolution operator on the triangle tensor $M$ will mess up the HSS picture, where the three-site triangle operator $I - \tau H_{\triangledown}$ is shown in (b).}
\label{fig:DistrHSS}
\end{figure}

First, as discussed in Sec. \ref{subsec:ral}, we expand the hexagon singlet with the over-complete basis of valence bonds as in Eq. \ref{eq:GDN} (with $\ket{\rm{D_-}}$ and $\ket{N}$ defined in Fig. \ref{fig:benzene}, see more details in Appendix \ref{variational}). 

As shown in Fig. \ref{fig:EsHSS}(a), by tuning $\omega$, we can indeed connect \textit{smoothly} the $\ket{\rm{BRS}}_O$ with the Hida point (at $\omega \simeq 0.5826$). Notably, the lowest energy state turns out to deviate slightly from the Hida point, although the energy difference between them is rather tiny.  {In addition, the exact expansion (\ref{eq:GDN}) at the Hida point means that Hida's HSS state, like the BRS in Fig. \ref{fig:ral}(a), also has a simple resonating AKLT-loop picture, but with both NN and third-NN valence bonds (instead of only NN bonds) across each hexagon}. 

Second, {we tune the tensor elements of hexagon tensor $R$, and thus explore all states in the HSS family. After accounting for all symmetry constraints, these tensor elements can all be expressed in terms of only two independent parameters, say, $\alpha$ and $\beta$} In Fig. \ref{fig:EsHSS}(b), $e_0$ versus energy difference ($e_A-e_B$) on two kinds of triangles are shown for each curve with fixed $\alpha$ and varying $\beta$ parameters {(see Tabs. \ref{tab:R-tensor-BRS}, \ref{tab:R-tensor} in Appendix \ref{QSpace-Tensor} for the specific definition).} From Fig. \ref{fig:EsHSS}(b) we can see that the global minimum is located at $e_A-e_B=0$, with the lowest energy per site found as $e_0 = -1.36000$, again only slightly lower than the value $e_0=-1.35991$ of the Hida point.

Besides the above two simulations, the third approach we have adopted is to introduce second- ($J_2$) and third-NN ($J_3$) couplings in the hexagon [see Eq. (\ref{eq:hexagon})], so as to strengthen longer-range valence bonds. Through tuning $J_2$ and $J_3$ in (\ref{eq:hexagon}), we find the same lowest variational energy as the above two calculations (not shown). Therefore, we conclude that the best variational energy of the spin-1 KHAF model, within the HSS states, is $e_0 = -1.36000$, located at $e_A-e_B=0$, i.e., without any trimerization order.

\begin{figure}
\includegraphics[width=1\linewidth]{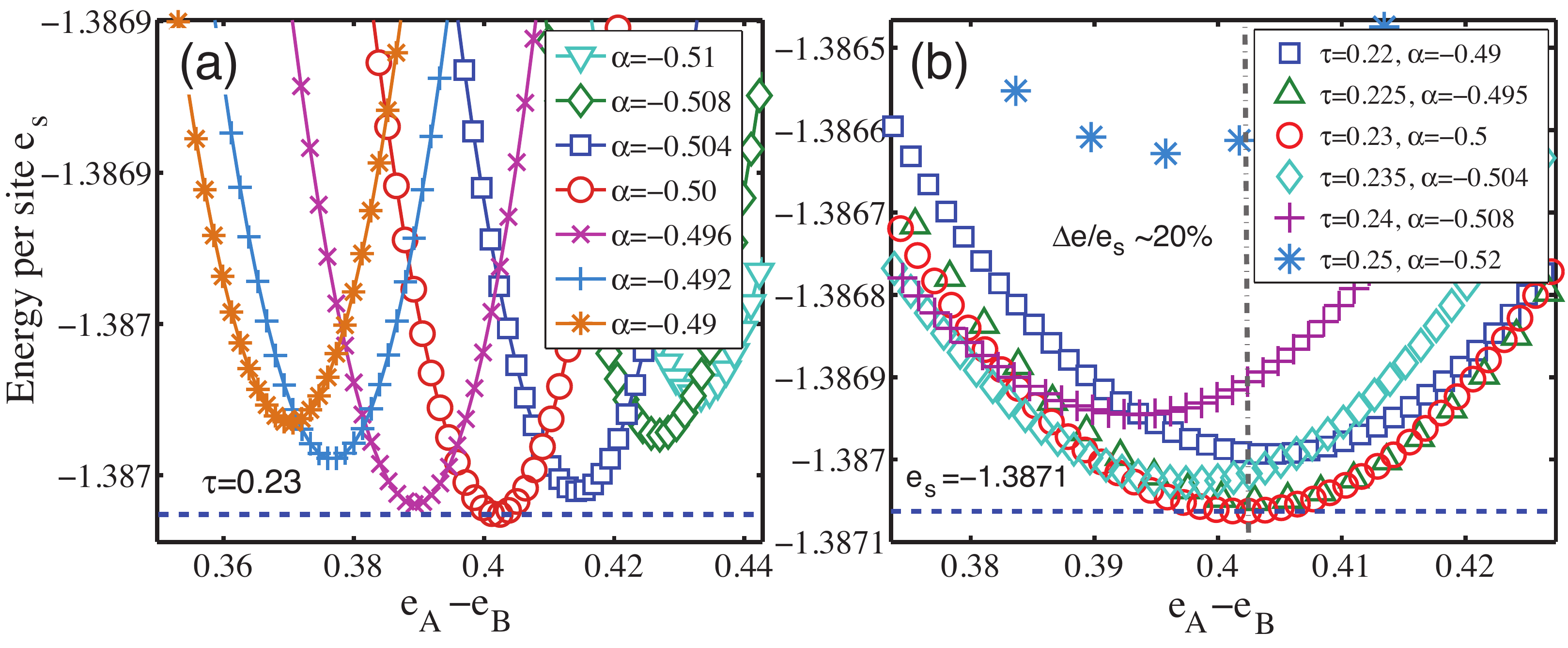}
\caption{(Color online)  (a) Energies per site $e_0$ for fixed $\tau=0.23$. By tuning $\alpha, \beta$, we find the lowest variational energy $e_0=-1.3871$ is at $\alpha=-0.5$ and $\beta=0.384$, with $e_A-e_B\neq0$. (b) We collect the curves which possess the lowest variational energy for various fixed $\tau$. The global minumn is on the $\alpha=-0.5$ and $\tau=0.23$ curve, and with $\Delta e/e_0\sim20 \%$ [$\Delta e = 2(e_A-e_B)/3$].}
\label{fig:EsTau}
\end{figure}

Beyond the HSS states, we are also interested in improving the variational energy of the spin-1 KHAF, within the present $D^*=2$ (i.e., two bond multiplets $0\oplus1$) tensor network ansatz shown in Fig.~\ref{fig:tensor}. Notice that the present $M$ (obtained by combining three projectors $P$) itself leaves the three physical spins uncorrelated, the correlation only enters
through the singlet construction in $R$. If we now instead introduce correlations directly between physical spins within $M$ across
hexagons (where $R$ still allows singlets within the hexagons only), the HSS picture will generally be \textit{shuffled} [see Fig. \ref{fig:DistrHSS}(a)]. In practice, we add one more parameter $\tau$, in addition to $\alpha$ and $\beta$, to tune tensor $M$ and thus explore the variational energies. 

In Fig. \ref{fig:EsTau}(a), inspired from the more sophisticated imaginary-time evolution approach, \cite{Vidal-2007, Orus-2008, Jiang-2008, Liu-2014a, Xie-2014} we show that when fixing the parameter $\tau \neq 0$, and scanning through various $\alpha, \beta$ parameters, the lowest energy state turns out to be located at $e_A-e_B \neq 0$; while the sign and magnitude of $\Delta e=2(e_A-e_B)/3$ depends on the values of the three parameters of $\alpha, \beta$ and $\tau$. We repeat the analysis of Fig.~\ref{fig:EsTau}(a) for various $\tau$-values and for each collect that curve containing the lowest variational energies, like the $\alpha=0.5$ curve for $\tau=0.23$ in Fig. \ref{fig:EsTau}(a). We show them in Fig. \ref{fig:EsTau}(b), from which it is found that the best
variational energy is $e_0=-1.3871$, significantly lower than the non-symmetry-breaking HSS state, and the energy difference $\Delta e = 2(e_A-e_B)/3 \approx 0.2683$, about $20\%$ of $e_0$. This implies that the SVBC has lower energy than any HSS states with $D^*=2$. This conclusion and even the magnitude of trimerization order parameter, obtained only by considering three parameters here, are in nice agreement with previous calculations of Refs. \onlinecite{Lauchli-2014, Liu-2014, Poilblanc-2014}, where the estimate $e_0 \approx -1.41$ of a SVBC state was obtained with much larger bond dimensions and more sophisticated numerical algorithms. However, our conclusion that SVBC states yield a lower energy than HSS states disagrees with the main conclusion of NN14 in Ref.~\onlinecite{Nishimoto-2014}. Possible reasons for this disagreement are given in Appendix~\ref{app:critique}.

\section{Summary and Discussion}
\label{sec:conclusion}
To summarize, we have performed a systematic investigation of the hexagon-singlet solid states for the spin-1 kagome antiferromagnet. Through the Schwinger boson representation, we have shown that the HSS states are gapped paramagnets, which have a hidden resonating AKLT-loop picture when the hexagon singlet is decomposed within the over-complete valence-bond bases. However, in contrast to the RAL state (equal weight superposition of all possible loop configurations) which is a $\mathbb{Z}_2$ spin liquid, the HSS states, owing to the local constraint of hexagon-singlet construction, turn out to have no intrinsic topological order. By performing numeric simulations using the tensor network representation, we have shown that the HSS states are indeed gapped spin liquid, with all correlation functions decaying exponentially, and the results of the entanglement entropy and spectra confirm that these states are topologically trivial. Furthermore, we find out that the best variational energy for the spin-1 KHAF model, among all the non-symmetry-breaking HSS states, is $e_0 = -1.36000$. Moreover, by an enlightening variational study, we have shown that, within the present $D^*=2$ tensor network ansatz, the simplex valence bond crystal state ($e_0=-1.3871$) is more energetically favorable than the non-symmetry-breaking HSS state.

An interesting issue we leave for a future study is to find a realistic Hamiltonian which could stabilize the HSS phase; such a Hamiltonian might contain second- or even third-NN couplings in the hexagons, which can still be conveniently treated in the present tensor network ansatz in Fig. \ref{fig:tensor}. Moreover, through the investigations of the RAL \cite{RAL-2014} and HSS states, it has been shown that the resonating AKLT loops, which constitute a rather general representation of spin-1 many-body singlets, are able to describe a variety of states, ranging from the topologically ordered RAL to the topologically trivial HSS states. Therefore, we expect that there could be more exotic quantum states by exploring the resonating AKLT loop family in future studies.

\section{Acknowledgment}
HHT acknowledges Tai-Kai Ng for introducing him the hexagon-singlet ansatz and Yi Zhou for stimulating discussions. WL and HHT would like to thank Meng Cheng and Zheng-Xin Liu for fruitful discussions on spin-1 kagome antiferromagnets. WL is further indebted to Tao Liu and Gang Su for helpful discussions. This work has been supported by the DFG project SFB-TR12, the EU project SIQS, and the DFG Cluster of Excellence NIM. AW further acknowledges support by DFG grant WE-4819/1-1.

\begin{appendix}
\setcounter{equation}{0}
\renewcommand{\theequation}{S\arabic{equation}}
\setcounter{figure}{0}
\setcounter{table}{0}
\renewcommand{\thefigure}{A\arabic{figure}}

\section{Exponentially decaying correlations in HSS ansatz}
\label{app:hss}

In this Appendix, we provide technical details on showing that the HSS ansatz all have exponentially decaying spin correlations. This relies on the spin-coherent state representation of the SU(2) Schwinger boson states. Following Refs. \onlinecite{Auerbach-book} and \onlinecite{Arovas-1988}, $S=1$ spin coherent states are defined by
\begin{equation}
|\vec{\Omega} \rangle =\frac{1}{2}(ub_{\uparrow }^{\dagger }+vb_{\downarrow
}^{\dagger })^{2}|\mathrm{vac}\rangle,
\end{equation}
where the O(3) vector $\vec{\Omega}$ is parametrized by the solid angle, $%
\vec{\Omega} =(\sin \theta \cos \phi ,\sin \theta \sin \phi ,\cos \theta )$,
and $u$ and $v$ are given by $u(\theta,\phi)=\cos \frac{\theta }{2}e^{-i%
\frac{1}{2}\phi }$ and $v(\theta ,\phi )=\sin \frac{\theta }{2}e^{i\frac{1}{2%
}\phi }$. The spin coherent states satisfy the following relations:
\begin{eqnarray}
\frac{3}{4\pi }\int d\vec{\Omega} \text{ }|\vec{\Omega} \rangle \langle \vec{%
\Omega} | &=&I, \\
\frac{3}{2\pi }\int d\vec{\Omega} \text{ }\vec{\Omega}|\vec{\Omega} \rangle
\langle \vec{\Omega} | &=& \vec{S},
\end{eqnarray}
where the integration $\int d\vec{\Omega}$ is over the solid angle, $\int d%
\vec{\Omega} =\int_{0}^{\pi }\sin \theta d\theta \int_{0}^{2\pi}d\phi$.

By using these relations, the norm of the HSS wave function (\ref{eq:HSS})
is expressed as
\begin{eqnarray}
\langle \Psi |\Psi \rangle  &=&\frac{3}{4\pi }\int d\vec{\Omega}\text{ }%
\langle \Psi |\vec{\Omega}\rangle \langle \vec{\Omega}|\Psi \rangle   \notag
\\
&=&\frac{3}{4\pi }\int d\vec{\Omega}\text{ }|\Psi (\vec{\Omega})|^{2},
\label{eq:norm}
\end{eqnarray}

\begin{widetext}
where $\Psi (\vec{\Omega})$ is given by
\begin{eqnarray}
\Psi (\vec{\Omega}) &=&\langle \Psi |\vec{\Omega}\rangle   \notag =\langle \mathrm{vac}|\prod_{\hexagon}\left[ \sum_{\{ij,kl,mn\}}\bar{w}%
_{ij}\bar{w}_{kl}\bar{w}_{mn}S_{\hexagon,ij}^{-}S_{\hexagon,kl}^{-}S_{%
\hexagon,mn}^{-}\right] \prod_{i}|\vec{\Omega}_{i}\rangle   \notag \\
&\propto &\prod_{\hexagon}\left[ \sum_{\{ij,kl,mn\}}\bar{w}_{ij}\bar{w}_{kl}%
\bar{w}_{mn}(u_{\hexagon,i}v_{\hexagon,j}-v_{\hexagon,i}u_{\hexagon,j})(u_{%
\hexagon,k}v_{\hexagon,l}-v_{\hexagon,k}u_{\hexagon,l})(u_{\hexagon,m}v_{%
\hexagon,n}-v_{\hexagon,m}u_{\hexagon,n})\right] ,
\end{eqnarray}
and $|\Psi (\vec{\Omega})|^{2}$ is written as
\begin{equation}
|\Psi (\vec{\Omega})|^{2}\propto \prod_{\hexagon}\left\vert
\sum_{\{ij,kl,mn\}}\bar{w}_{ij}\bar{w}_{kl}\bar{w}_{mn}(u_{\hexagon,i}v_{%
\hexagon,j}-v_{\hexagon,i}u_{\hexagon,j})(u_{\hexagon,k}v_{\hexagon,l}-v_{%
\hexagon,k}u_{\hexagon,l})(u_{\hexagon,m}v_{\hexagon,n}-v_{\hexagon,m}u_{%
\hexagon,n}) \right\vert ^{2}.
\end{equation}%

Note that the norm $\langle \Psi |\Psi \rangle $ in (\ref{eq:norm}) is proportional to the following partition function of a classical statistical model defined on the same kagome lattice:
\begin{eqnarray}
Z &=&\int d\vec{\Omega}\text{ }|\Psi (\vec{\Omega})|^{2}  \notag \\
&=&\int d\vec{\Omega}\prod_{\hexagon}\left\vert \sum_{\{ij,kl,mn\}}%
\bar{w}_{ij}\bar{w}_{kl}\bar{w}_{mn}(u_{\hexagon,i}v_{\hexagon,j}-v_{\hexagon%
,i}u_{\hexagon,j})(u_{\hexagon,k}v_{\hexagon,l}-v_{\hexagon,k}u_{\hexagon%
,l})(u_{\hexagon,m}v_{\hexagon,n}-v_{\hexagon,m}u_{\hexagon,n})
\right\vert ^{2}  \notag \\
&=&\int d\vec{\Omega}\exp \left( \ln \prod_{\hexagon}\left\vert %
 \sum_{\{ij,kl,mn\}}\bar{w}_{ij}\bar{w}_{kl}\bar{w}_{mn}(u_{\hexagon%
,i}v_{\hexagon,j}-v_{\hexagon,i}u_{\hexagon,j})(u_{\hexagon,k}v_{\hexagon%
,l}-v_{\hexagon,k}u_{\hexagon,l})(u_{\hexagon,m}v_{\hexagon,n}-v_{\hexagon%
,m}u_{\hexagon,n}) \right\vert ^{2}\right)   \notag \\
&=&\int d\vec{\Omega}\exp \left( 2\sum_{\hexagon}\ln \left\vert %
 \sum_{\{ij,kl,mn\}}\bar{w}_{ij}\bar{w}_{kl}\bar{w}_{mn}(u_{\hexagon%
,i}v_{\hexagon,j}-v_{\hexagon,i}u_{\hexagon,j})(u_{\hexagon,k}v_{\hexagon%
,l}-v_{\hexagon,k}u_{\hexagon,l})(u_{\hexagon,m}v_{\hexagon,n}-v_{\hexagon%
,m}u_{\hexagon,n}) \right\vert \right)   \notag \\
&=&\int d\vec{\Omega}\exp \left( -\sum_{\hexagon}h_{\hexagon}\right),
\end{eqnarray}%
where $h_{\hexagon}$ is the classical Hamiltonian describing local interactions among the O(3) vectors within each hexagon%
\begin{equation}
h_{\hexagon}=-2\ln \left\vert \sum_{\{ij,kl,mn\}}\bar{w}_{ij}\bar{w}_{kl}\bar{w}_{mn}(u_{\hexagon,i}v_{\hexagon,j}-v_{\hexagon,i}u_{\hexagon
,j})(u_{\hexagon,k}v_{\hexagon,l}-v_{\hexagon,k}u_{\hexagon,l})(u_{\hexagon,m}v_{\hexagon,n}-v_{\hexagon,m}u_{\hexagon,n}) \right\vert.
\end{equation}

\begin{figure}
\includegraphics[width=0.85\linewidth]{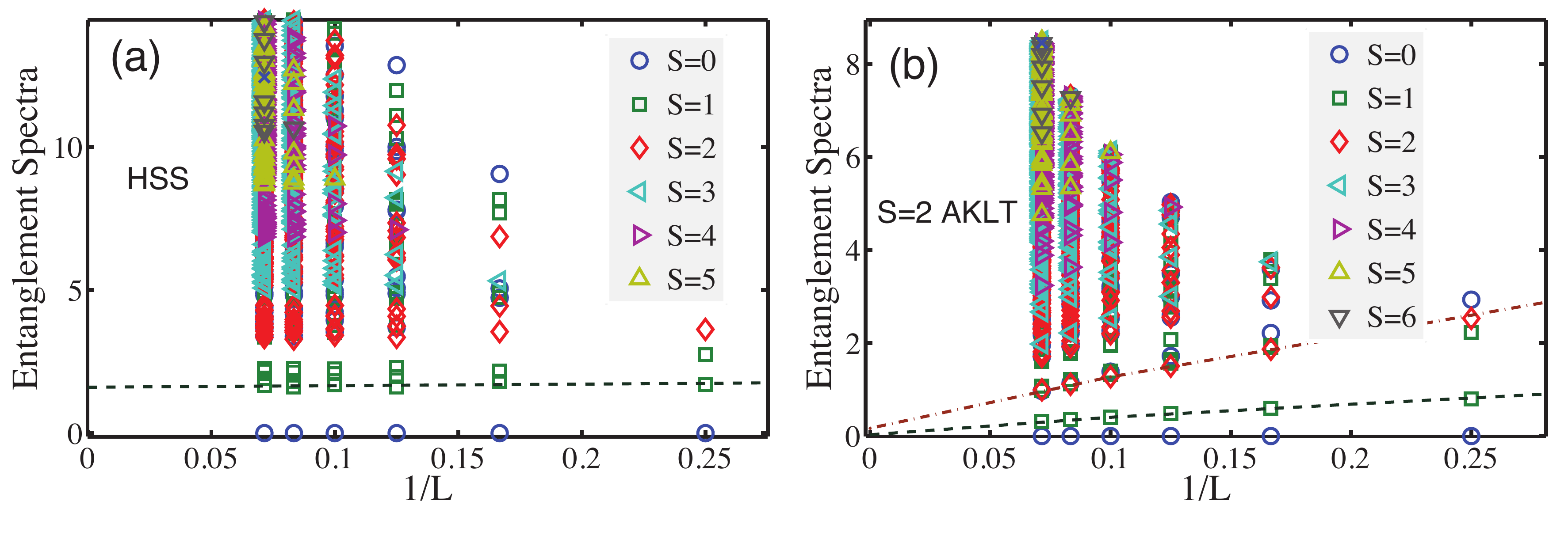}
\caption{(Color online) The entanglement spectra (ES) on YC geometries, with various cylinder widths $L$. Each symbol in the figure labels a multiplet, instead of an individual plain state. The spectra of different circumferences have been offset so as the lowest eigenvalues are zero. (a) The ES of Hida's HSS state on cylinders with $L=4,6,8,10,12,14$. There exists a non-vanishing singlet-triplet gap ($\delta \sim1.6$) in the spectra. (b) The ES of the $S=2$ square-lattice AKLT state on cylinders with $L=4,6,8,10,12,14$, where the singlet-triplet gaps $\delta$ ($\sim 1/L$) extrapolate to zero in the large $L$ limit.}
\label{fig:ES}
\end{figure}

\end{widetext}

Similarly, the spin-spin correlation function can be represented by using spin coherent states as
\begin{eqnarray}
\frac{\langle \Psi |\vec{S}_{i}\cdot \vec{S}_{j}|\Psi \rangle }{\langle \Psi
|\Psi \rangle } &=&\frac{4}{Z}\int d\vec{\Omega}\text{ }|\Psi (\vec{\Omega}
)|^{2}\vec{\Omega}_{i}\cdot \vec{\Omega}_{j}\text{ \ \ \ \ \ }(i\neq j)
\notag \\
&=&\frac{4}{Z}\int d\vec{\Omega}\text{ }\vec{\Omega}_{i}\cdot \vec{\Omega}
_{j}\exp \left( -\sum_{\hexagon}h_{\hexagon}\right) .  \label{eq:correlator}
\end{eqnarray}
This shows that the spin-spin correlation function in the HSS ansatz is equivalent to the two-point correlation function between the O(3) vectors in a 2D classicial statistical model. Since the classical statistical model is at finite temperature and has only short-range interactions, Mermin-Wagner theorem\cite{Mermin-1966} indicates that there is no long-range order and the correlation function (\ref{eq:correlator}) decays exponentially.

{Notice, however, that there still exists the possibility that the classical Hamiltonian in Eq. (S8) is unbounded for certain choices of $w_{ij}$ (due to the interaction in a logarithmic form). In that case, the Mermin-Wagner theorem would not apply directly and a more rigorous approach would be needed. In any case, our numerical evidence strongly suggests that our argument does hold for general $w_{ij}$.}

Other correlation functions, such as the quadrupole-quadrupole and dimer-dimer correlation functions, can be similarly represented as correlation functions in the same classical statistical model. Apart from a qualitative understanding of their behaviors using the Mermin-Wagner theorem, an additional benefit of this quantum-classical mapping is that Monte Carlo techniques can also be applied directly to compute the physical quantities accurately (see e.g. Ref. \onlinecite{Sondhi-2009} for this application to the 3D AKLT states). 

{\section{The Hida state as a resonating AKLT-loop state}
\label{variational}
In this appendix, we discuss the expansion of Hida's HSS state with the valence bond basis and thus reveal that it belongs to the family of resonating AKLT-loop state (RAL), therefore elaborating the discussion in Section \ref{subsec:ral}. By comparing the tensor elements in Tabs. \ref{tab:R-tensor-BRS} and \ref{tab:R-tensor} in Appendix \ref{QSpace-Tensor}, we can see that Hida's HSS state has similar weights as $\ket{\rm{BRS}}_O$ in corresponding channels, which suggests that these two states may have a big overlap, and the dominating hexagon-singlet configuration in the former might consist of NN valence bonds. Thus we could add hexagon-singlet configuration containing longer-range bonds {in addition to the configurations $\ket{D_-}$} of the $\ket{\rm{BRS}}_O$ [Fig. \ref{fig:benzene}(a)]. Here we consider the configuration in Fig.~\ref{fig:benzene}(b), which consists of two NN and one second NN bonds in each hexagon. Associating a weight $\omega$ with this hexagon-singlet configuration (denoted as $\ket{N}$), i.e., the singlet state can be defined as $\ket{\psi_{\hexagon}(\omega)} = \ket{\rm{D_-}}+ \omega \ket{N}$ in each hexagon, and we thus construct a one-parameter HSS state family $\ket{\Psi(\omega)} = \bigotimes_i P_i \prod_{\hexagon} \ket{\psi_{\hexagon}(\omega)}$ according to Eqs. (\ref{eq:psi}-\ref{eq:projector}) in the main text. 

At a first glance, the fact that the hexagon singlet $\ket{\rm{G}}$ (i.e., the ground state of a six-site Heisenberg ring) in Hida's HSS state can be expressed exactly as a superposition of $\ket{\rm{D}_-}$ and $\ket{\rm{N}}$ at $\omega=0.582618977$ (see Fig. \ref{fig:EsHSS} in the main text) is surprising, since chances are these three normalized while non-orthogonal states (vectors) do not lie on the same plane. In fact, there are two independent components $\alpha$ and $\beta$ for the $C_3$ lattice rotational invariant states, as shown in Tabs. \ref{tab:R-tensor-BRS} and \ref{tab:R-tensor}, which can parametrize a sphere (after proper reorganizations). Nevertheless, it turns out that these three states satisfy the condition $\langle\rm{G}|D_-\rangle^2 + \langle\rm{G}|N\rangle^2 + \langle\rm{N}|D_-\rangle^2 = 1+ 2 \langle\rm{G}|D_-\rangle \langle\rm{G}|N\rangle  \langle\rm{N}|D_-\rangle$, meaning $\ket{\rm{G}}$, $\ket{\rm{N}}$, and $\ket{\rm{D}_-}$ are on the same unit circle and thus are linearly dependent. As a more careful analysis shows, the reason for this is that these hexagon singlets (and the corresponding non-symmetry-breaking HSS states) are {all 1D irreducible representations of the full $C_{6v}$ point group. For example, they are} eigenstates of the hexagon-inversion symmetry operator (with eigenvalues $\pm1$). Therefore, symmetry adds one more constraint and leaves only one independent parameter $\omega$ in Eq. (\ref{eq:GDN}) when one expands the hexagon singlet state $\ket{\rm{G}}$ with $\ket{\rm{N}}$ and $\ket{D_-}$ of the same inversion parity ($-1$), the other state $\ket{\rm{D_+}}$ in Fig. \ref{fig:benzene} not used in the expansion of $\ket{\rm{G}}$ above belongs to an $+1$ eigenstate of hexagon-inversion symmetry).
} 

{\section{The QSpace representation of the HSS states}
\label{QSpace-Tensor}
In this appendix, we show the specific SU(2)-invariant tensor-network representations of the HSS states, including the BRS and Hida's HSS state, thereby elaborating the discussion in Section \ref{sec:tensor}.

For the BRS's, corresponding $R$-tensors are shown in Tab. \ref{tab:R-tensor-BRS}, where $R$ is described in the QSpace language, a practical framework for implementing non-abelian symmetries in tensor networks.\cite{AWb-2012} {In the QSpace framework, the tensor $R$ can be decomposed into the reduced multiplet data $||R||$ and the Clebsch-Gordan coefficients (CGCs) $C$, i.e.,
\begin{equation}
R = \sum_{\{S_i\}} \sum_{\{m_i^z\}} ||R||_{S_1, S_2, S_3}
(C_{S_1,S_2,S_3}^{{S_{\rm hex}} = 0})_{m_1^z, m_2^z, m_3^z},
\end{equation}
where $\{S_i\}$ are the spin quantum numbers and $m_i \in [-S_i, +S_i]$ are the $z$ component of magnetic quantum number of paired-up virtual spins.} Notice that $R$ represents a tensor of order four where the fourth dimension specifies the symmetry of the total tensor, which here for the HSS state by definition always a singlet state {$(S_{\rm hex}=0)$}. The last dimension therefore represents a singular dimension which can be safely ignored. Overall then, what represents a $4\times4\times4$ tensor in full state space becomes a $2\times2\times2$ tensor in the reduced multiplet space. 

In Tab. \ref{tab:R-tensor-BRS}, the first column enumerates different records or fusion channels, i.e., the different ways three virtual particles can be {coupled into a hexagon spin singlet}. The second column shows the spin quantum numbers (S-labels) {of these virtual particles}: Respectively, there are three spin-0 (in record No. 1), two spin-1's and a spin-0 (No. 2-4), and three spin-1 particles (No. 5). {The third column contains the dimensions of the CGC tensor $(C_{S_1,S_2,S_3}^{q=0})_{m_1^z, m_2^z, m_3^z}$, which vary in different records because of different spin quantum numbers $[S_1,S_2,S_3]$.}  {The last column gives the value of the reduced tensor element obtained when coupling $S_1$, $S_2$ and $S_3$ together to obtain a hexagon singlet.} The norm of $||R||$ is chosen such that the weight in the first record is 1, the other weights are denoted as $\alpha$ in records 2-4, and $\beta$ in the last record, respectively. Note that in Tab. \ref{tab:R-tensor-BRS} and other tables for QSpace tensors hereafter, the actual number of the CGC coefficients are not shown. In practice, the CGC tensor in each channel are normalized in such a way that the largest tensor element equals $\pm1$, and the first nonzero coefficient being positive.

Similar to the hexagon tensor $R$, the triangle tensor $M$ can also be determined by performing a few steps of contractions [as indicated in Fig. \ref{fig:tensor}(d)], with the three physical $S=1$ spins combined into a single state space ($3^3=27$ states reduced to 7 multiplets), and this results in a tensor of rank four. 

\begin{table}[h]
\centering
\caption{Hexagon tensor $R$ of even and odd BRS, as a QSpace object. {$\{S_i\}$ are the spin quantum numbers, and the third column after it demonstrates the dimensions of corresponding CGC tensor in each channel (simply also determined from the symmetry labels $S_i$), the fourth column stores the reduced multiplet elements $||R||$. {The values for $\alpha$ and $\beta$ listed here define the even and odd BRS state, discussed in Section \ref{sec:tensor} A. In Section \ref{sec:variational}, though, they are used as variational parameters. }}}\label{tab:R-tensor-BRS}
\begin{tabular}{ c c c c c }
  \hline \hline
  $|\rm{BRS}\rangle_E$ &  $[S_1,S_2,S_3]$ & dimensions & $||R||$ \\
  1.  & [  0 ,  0 ,  0  ] & 1$\times$1$\times$1 & 1 \\
  2. & [  0 ,  1 ,  1  ] & 1$\times$3$\times$3 & $\alpha=1/3$ \\
  3. & [  1 ,  0 ,  1  ] & 3$\times$1$\times$3 & $\alpha=1/3$ \\
  4. & [  1 ,  1 ,  0  ] & 3$\times$3$\times$1 & $\alpha=1/3$ \\
  5. & [  1 ,  1 ,  1  ] & 3$\times$3$\times$3 & $\beta=-1/3$ \\

\hline
  $|\rm{BRS}\rangle_O$ &   &  &  \\
  1.  & [  0 ,  0 ,  0 ] &  1$\times$1$\times$1 & 1 \\
  2. & [  0 ,  1 ,  1  ] &  1$\times$3$\times$3 & $\alpha=-0.2$ \\
  3.  & [  1 ,  0 ,  1  ] &  3$\times$1$\times$3 & $\alpha=-0.2$ \\
  4. & [  1 ,  1 ,  0  ] & 3$\times$3$\times$1  & $\alpha=-0.2$ \\
  5.  & [  1 ,  1 ,  1  ] &  3$\times$3$\times$3 & $\beta=0.2$ \\
  \hline \hline
\end{tabular}
\end{table}

On the other hand, for Hida's HSS state, the QSpace representation of tensor $R$ is shown in Tab. \ref{tab:R-tensor}, which is the lowest singlet ground state of a hexagonal Heisenberg ring. Since there are only two multiplets (a $S=0$ singlet and a $S=1$ triplet) on each geometric bond, the reduced bond dimension is thus again $D^*=2$ (i.e., two multiplets), corresponding to four individual states ($D=4$).}

\begin{table}[h]
\centering
\caption{Hexagon tensor $R$ of Hida's HSS state. Notice that to keep the $C_3$ discrete rotational lattice symmetry, the coefficients of records 2 to 4 should be equal (denoted by $\alpha$).}\label{tab:R-tensor}
\begin{tabular}{ c c c c c }
\hline
  No. & $[S_1, S_2, S_3]$ & dimensions & $||R||$ \\
  1. & [  0,  0,  0 ] & 1$\times$1$\times$1 & 1 \\
  2. & [  0,  1,  1 ] & 1$\times$3$\times$3 & $\alpha=-0.2457$ \\
  3. & [  1,  0,  1 ] & 3$\times$1$\times$3 & $\alpha=-0.2457$ \\
  4. & [  1,  1,  0 ] & 3$\times$3$\times$1 & $\alpha=-0.2457$ \\
  5. & [  1,  1,  1 ] & 3$\times$3$\times$3 & $\beta=0.1315$ \\
\hline
\end{tabular}
\end{table}

\section{Entanglement Spectra}
\label{app:spectra}

In this Appendix, we show the results of the entanglement spectra (ES) on cylinders with various widths $L$. By performing exact contraction for small $L$ and MPS-based approximate contractions for large $L$ ($\geq12$), one can get converged left and right vectors $V_l(i, i')$ and $V_r(i, i')$ after a sufficient number (typically $10\sim20$) of iterations. The vectors $V_l, V_r$ are the dominating (left and right) eigenstates of the transfer-matrix of the double-layer cylinder tensor network; and $|i), |i')$ are the virtual bases on the geometric bond. Since the $|i)$ and $|i')$ bases are not orthonormal to each other, it is important to notice that $V_l, V_r$ themselves are not the reduced density matrices of the half-cylinder. The reduced density matrix can be obtained through transformation $\rho = \sqrt{V_r} V_l \sqrt{V_r}$.\cite{Cirac-2011}

By diagonalizing $\rho = U \, \Lambda \, U'$, in Fig. \ref{fig:ES} we
show the entanglement spectrum, i.e., the minus logarithmic of the
density matrix spectrum $-\log(\Lambda)$, on cylinders of various
widths. Notice that since SU(2) symmetry has been implemented in the
tensor network, each symbol in Fig. \ref{fig:ES} represents a
multiplet with well-defined spin quantum bumber $S$, so we can tell
the spin $S$ of each level. In Fig. \ref{fig:ES}(a), there is a
distinct gap between lowest singlet and the first excited triplet
state, which does \textit{not} vanish as $L$ increases. This is in
contrast to a typical $S=2$ AKLT state on a square lattice: in
Fig. \ref{fig:ES}(b) we show the ES of the $S=2$ AKLT state on
cylinders, for which the singlet-triplet gap $\delta$ is propositional
to the inverse width $1/L$. $\delta$ vanishes in the infinite-$L$
limit, suggesting a gapless edge mode, in agreement with the results
in Refs. \onlinecite{Cirac-2011, Lou-2011}. The different behaviors of
the HSS and $S=2$ square-lattice AKLT states implies that the former
is a trivial insulator phase, and does not belong to a SPT phase.

\section{Comments on NN14's DMRG study on HSS states}
\label{app:critique}

{In this Appendix, we {address} the relation between our present
  tensor network study of the HSS state and {the work of NN14 in
    Ref. \onlinecite{Nishimoto-2014}, who studied the same model using
    DMRG, but, in contrast to us and
    Refs.~\onlinecite{Lauchli-2014,Liu-2014,Poilblanc-2014}, concluded that
    the model's ground state is an HSS state. They performed} DMRG
  simulations for the spin-1 KHAF on four types of clusters: (i)
  cylindrical clusters; (ii) periodic clusters; (iii) clusters with
  open boundary conditions (OBC) purposefully designed to favor a HSS
  ground state, by choosing cluster shapes that contain only hexagons
  around the edges and putting spin-1/2's on the outermost sites
  {(called HSS clusters)}; and (iv) clusters with OBC purposefully
  designed to favor SVBC order, by choosing cluster shapes that
  contain only triangles around the edges (called SVBC clusters).  For
  (i) to (iii) they found ground states without clear signatures of
  trimerization, while for (iv) they found a ground state that clearly
  trimerizes. Their conclusion of vanishing trimerization order on (i)
  is in direct contradiction to CL's observation that stable
  trimerization order exists on cylinders with width
  $L=8$.\cite{Lauchli-2014} NN14 estimated the energy per
  site in the bulk, $e_0$, by finite-size extrapolations to the
  thermodynamic limit. Their SVBC clusters, (iv), yielded the highest
  extrapolated energy ($e_0 = -1.391(2)$); their cylindrical and
  periodic clusters, (i) and (ii), yielded extrapolated energies ($e_0
  = -1.40988$ and $e_0 = -1.409(5)$, respectively) consistent with
  those of CL\cite{Lauchli-2014} and Liu \textit{et al.}
  \cite{Liu-2014}; and their HSS clusters, (iii), yielded the lowest
  extrapolated energy ($e_0 = -1.41095$). Based on this evidence, NN14
  argued that the ground state is a HSS state, and not the SVBC state
  advocated in Refs.~\onlinecite{Lauchli-2014,Liu-2014,Poilblanc-2014}.}

In our opinion, NN14's conclusion is flawed because {the tool that they use to distinguish the HSS and SVBC states, namely ground-state DMRG on finite-size clusters, relies on boundary effects to distinguish these states.}  However, such a tool can not reliably
estimate the difference in \textit{bulk} $e_0$ for these two types of state, since by definition, the \textit{bulk} value of $e_0$ is the value obtained for clusters so large that boundary effects vanish. Not surprisingly, the data for their HSS and SVBC clusters shown in their Fig.~4 have a much stronger finite-size dependence (larger slope of data plotted versus inverse system size) than their cylinder and periodic clusters. The fact that their extrapolations from the HSS and SVBC clusters disagree with each other in the thermodynamic limit just means that these clusters are not yet large enough to give reliable bulk $e_0$ values. Thus the discrepancy between NN14's $e_0$ values for HSS and SVBC clusters in our opinion does not reflect the true difference in the $e_0$ values of bulk SVBC and bulk HSS states; rather, it reflects the error bar, induced by boundary effects, in their determination of the bulk $e_0$ of the true ground state. 

{In addition, we also would like to point out that in Ref. \onlinecite{Ganesh-2013}, a strategy similar to that of NN14 has been applied to $J_1$-$J_2$ honeycomb Heisenberg model, there, too, inconsistent results were reported when changing the cluster shapes.}

{We would also like to make a comment regarding the trick of pinning
spin-1/2's on the open boundary of a spin-1 model. This trick was}
first developed for simulating a spin-1 Heisenberg chain, where two
spin-1/2's are put on the ends of the open chain to remove the edge
states (and related degeneracies).\cite{White-1993} {Some of us
  (WL, AW, JvD) have actually used this trick ourselves in recent work
  on spin-1 Heisenberg chains.\cite{Li-2013}} In that context, this
trick leads to a better numerical convergence for the bulk properties,
since it binds the edge modes localized on both open
ends. Importantly, however, it does \textit{not} change the bulk
physics, i.e., the \textit{same} value for $e_0$ is obtained with or
without employing spin-1/2's on the ends. \cite{White-1993,
  Li-2013} In contrast, in NN14's work, the presence or absence of
spin-1/2 particles on the outermost sites of their HSS or SVBC
cluster, respectively, is an attempt to use boundary effects to
stabilize two states with \textit{different} bulk properties. As argued
above, this strategy is intrinsically flawed. Thus, in our opinion
NN14 are mistaken in believing that their DMRG calculations on HSS
clusters can be used to {reliably} predict the bulk properties of
HSS states. Their DMRG calculations target purely ground state
properties, and if a certain type of state (here the HSS state)
happens \textit{not} to be the actual ground state (as argued in the
present paper and in Refs. \onlinecite{Lauchli-2014, Liu-2014,
  Poilblanc-2014}), then its bulk properties are simply inaccessible
to NN14's DMRG simulations. In contrast, through the present work we find an
explicit tensor network state representation for the HSS state family,
which makes Hida's HSS state readily accessible with tensor network
methods. We show unambiguously that the best HSS variational energy
for the spin-1 KHAF model is $-1.3600$, much higher than NN's ``HSS"
energy ($-1.41095$).

{In contrast to NN14's HSS and SVBC data, their data for cylinder
and periodic clusters show much weaker finite-size effects (smaller
slopes when plotted versus inverse system size). We believe that the
$e_0$ values extrapolated from this data \textit{do} reflect the
bulk $e_0$ value of the true ground state rather accurately -- they
are certainly consistent with our own $e_0$ values and those of
CL. We find it all the more surprising and perplexing that} NN14 did
not observe evidence for SVBC order in their cylinder clusters,
whereas CL, who used essentially similar cylindrical clusters, did.\cite{Lauchli-2014}
We suspect that in NN14's calculations the
presence of a symmetry-breaking ground state was still hidden by
boundary effects, which can cause a non-symmetry-breaking ground state
to appear through superposition. Indeed, CL reported that in their
calculations SVBC order emerged clearly only in their cylinders of
largest circumference (width 8).

{We conclude with a comment about the bond dimension used in the
present tensor network study, namely $D^* = 2$ (corresponding to
$0\oplus1$ in Tabs. \ref{tab:R-tensor-BRS} and \ref{tab:R-tensor}),
or $D=4$. One might ask: is our conclusion in
Section~\ref{sec:variational} that HSS states have higher energy
than SVBC states robust with respect to increasing $D$? Perhaps HSS
states would yield a lower energy than SVBS states if each hexagon
were allowed to involve not only spin-1/2 virtual particles but also
higher-spin virtual particles?} In this regard, we {note} that
the HSS states studied here are a subset of {the much more general
class of} tensor network states {studied by Liu \textit{et al.}}
in Ref.~\onlinecite{Liu-2014}, where tensor network variational
calculations were performed with bond dimensions as large as $D=20$,
implying {that} many more virtual particles with higher spin
{were included}. Those large-$D$ calculations showed unambiguously
that the spin-1 KHAF model possesses a SVBC ground state with
$e_0=-1.41035$. {Therefore, the conclusion of the present work,
that HSS states lie significantly higher in energy than SVBC states,
is not specific to using an HSS ansatz with $D=4$, as done here;
instead, the results of Ref.~\onlinecite{Liu-2014} demonstrate that
it holds throughout as $D$ is increased up to $D=20$, a value
sufficiently large that the results reported in
Ref.~\onlinecite{Liu-2014} were well converged.}

\end{appendix}
\twocolumngrid

\end{document}